\documentclass[11pt,A4]{article}
\usepackage{booktabs}

\usepackage{hyperref}
\usepackage{url}
\usepackage{setspace}
\usepackage[T1]{fontenc}
\usepackage{lscape}

\usepackage[svgnames]{xcolor}
\usepackage{amsmath,amsthm,amsfonts}
\usepackage{rotating}

\usepackage[round]{natbib}
\usepackage{xcolor}

\def\blue{\color{blue}}
\def\blue{}

\theoremstyle{plain}
\theoremstyle{remark}
\theoremstyle{definition}

\DeclareMathSymbol{\Re}{\mathalpha}{AMSb}{"52}

\newcommand{\iid}{\stackrel{\texttt{iid}}{\sim}}
\newcommand{\silvia}[1] {{\color{black}{#1}}}

\usepackage{comment, mathtools}
\MHInternalSyntaxOn
\newcommand{\adjintertext}[3]
{\ifvmode\else\\\@empty\fi
  \noalign{%
    \vskip-\lineskiplimit      
    \vskip\normallineskiplimit 
    \vskip#1
     \vbox{\normalbaselines
       \ifdim
         \ifdim\@totalleftmargin=\z@
           \linewidth
         \else
           -\maxdimen
         \fi
       =\columnwidth
      \else \parshape\@ne \@totalleftmargin \linewidth
      \fi
      \noindent#3\par}%
    \vskip-\lineskiplimit      
    \vskip\normallineskiplimit 
    \vskip#2
 }}%
\MHInternalSyntaxOff

\begin{document}

\title{A unit-level small area model with misclassified covariates}
\author{{\bf Serena Arima}\thanks{Corresponding author: Serena Arima Dipartimento di
metodi e modelli per l'economia, la finanza e il territorio,
Sapienza Universit\`a di Roma, Via del Castro Laurenziano, 9
00161 Roma, Italia; e-mail: \texttt{serena.arima@uniroma1.it}.
} \\
Sapienza Universit\`{a}\ di Roma \\
{\bf Silvia Polettini}\\ Sapienza Universit\`{a}\ di Roma}

\maketitle

{\bf Running headline:} Small area model with perturbed covariates

\begin{abstract}
Model-based small area estimation relies on mixed effects regression models that link the small 
areas and borrow strength from similar domains. 
When the auxiliary variables used in the models are measured with error,  small area estimators that 
ignore the measurement error may be worse than direct estimators. Alternative small area estimators accounting for measurement error have been proposed in the literature but only for continuous auxiliary variables. Adopting a Bayesian approach, we extend the unit-level model in order to account for measurement error in both continuous and categorical covariates. For the discrete variables we model the misclassification probabilities and estimate them jointly with all the unknown model parameters.  We test our model through a simulation study.
The impact of the proposed model is emphasized through application 
to data from the Ethiopia Demographic and Health Survey where we focus on the women's malnutrition issue, a dramatic problem in developing countries and  an important indicator of the socio-economic progress of  a country.
\end{abstract}


\section{Introduction}
\label{sec:00}
In survey sampling, small area estimation aims at estimating aggregates of interest over unplanned domains when the sample sizes are not sufficient to obtain reliable design based (direct) estimates.  Model based approaches to small area estimation  focus on mixed effects regression models that link the small 
areas and borrow strength from similar domains.  
However, it might be the case that the auxiliary variables used in such models are measured with error.  In regression models, 
the presence of measurement error in covariates is known to cause biases in estimated model parameters and lead to loss of power for detecting interesting relationships among variables \citep{Carroll:2006}.  In small area estimation, 
ignoring such error may produce estimators that perform worse than direct estimators \citep{ybarra:2008,Arima:2014}. 
Corrections to the unit-level and area-level models have been proposed both in a frequentist and a Bayesian context, but limited to the case of continuous covariates \citep{Ghosh:2006, Ghosh:2007,ybarra:2008, Datta:2010,Arima:2014}.  We discuss these issues in the context of unit-level small area models, when covariates subject to measurement error are of categorical nature. We propose a unit-level 
small area model able to deal with measurement error in categorical as well as continuous covariates.  A clear example of the effect of neglecting measurement error and an illustration of the advantages of the proposed procedure arise in the analysis of body mass index (BMI) of Ethiopian women, that we base on 2011 Ethiopia Demographic and Health Survey (DHS) data\footnote{Data are collected under the MEASURE DHS project, funded by United States Agency for International Development (USAID).}. BMI is taken as a measure of women's nutritional status, a key indicator of
%
%
%
the socio-economic development of a country. 
Not surprisingly, for many countries this aspect has been the object of prioritized interventions in the achievement of the Millennium Development Goals.
%
%
%
Although undernutrition has been reduced in recent years, yet, food insecurity remains the greatest challenge in Ethiopia and a serious drawback to the country's economic development; 
 moreover, high regional as well as socio-economic disparities remain. 
For the above reasons, it would be important to obtain accurate estimates of women's mean BMI levels across domains, first of all those defined by the administrative regions.  
The model allows investigating the role on BMI of a number of socio-economic characteristics such as age, household's wealth index, number of children, and  level of educational attainment, while accounting for regional variation, that is large in the country. 
%
%
All of the above variables are clearly potentially explicative of the woman's nutritional status and highlighted as important determinants of undernutrition in previous studies \citep{DHS}. However, for some of them it is reasonable to assume that they are measured with error. 
 Our application reveals that, even in the presence of large subsamples, the small area predictions obtained ignoring the measurement error may be misleading and covariates' effect  may be severely altered.

The paper is organized as follows: we start with a brief review of the literature on measurement error models in Section~\ref{sec:01} and focus on this issue in small area context (Section \ref{sec:02}). In Section~\ref{sec:03} we describe our proposal;  Section~\ref{sec:04} is devoted to a simulation study that investigates the performance of the proposed model. In Section~\ref{sec:05} the model is applied to the Ethiopia DHS data. We conclude with a discussion in Section~\ref{sec:06}.

\section{Measurement error models}
\label{sec:01}

Measurement error in covariates is an established and well known problem, and there is an enormous literature on this topic  \citep[see, among the others,][]{Fuller:1987,Carroll:2006}.
Although measurement error models have been mainly developed for the analysis of experimental data, their role in social studies and, in particular, in official statistics, is crucial. Modern  small area methods  heavily rely on the availability of good auxiliary information entering the model in the form of covariates.  Such covariates are often  estimates  obtained by a larger survey, administrative sources, or a previous census; sometimes they arise as the result of field measurement and lab analysis \citep{Ghosh:2006,Buonaccorsi:2010}.
 As a consequence, we do not observe the true level of the covariate, but only an estimate. 
Also, covariates may be self-reported responses \citep{ybarra:2008}, for which under-reporting,  lack of memory and digit preference may occur. Under these circumstances,  it may be assumed that covariates are measured with error. 

The presence of measurement error in covariates causes biases in estimated model parameters and  leads to loss of power for detecting interesting relationships among variables.



Most of the measurement error literature relies on the classical measurement error model \citep{Fuller:1987,Carroll:2006}. 
Following the notation of \citet{Ghosh:2006}, we denote by uppercase letters the variables observed with error, and by lowercase letters the corresponding latent values.
The measurement error model assumes  that for each single unit, the covariate $x_{i}$ $(i=1,...,n)$ is not available; instead, we observe $r \geq 1$ replications of measurements of $x_{i}$ subject to additive error, namely
\begin{equation}
\label{Meas-Err}
X_{ij}=x_{i} + \eta_{ij} \quad j=1,...,r
\end{equation}
where $\eta_{ij}$'s are independent and identically distributed variables with zero mean. The $x_{i}$'s might be either unknown, fixed, quantities or random variables. In the first case, the measurement error is called functional, whereas in the second  it is defined as structural. 

Model in equation (\ref{Meas-Err}) assumes that the mismeasured covariates are continuously distributed.
For discrete covariates, measurement error means misclassification. Examples abound: item preference for privacy or social desirability reasons, digit preference and recall errors are common sources of misclassification; the discretized version of a continuous covariate measured with error is also subject to misclassification. 
Also, misclassification error may be artificially induced for disclosure limitation purposes by National Statistical Offices \citep{PRAM, Polettini:2015}.

In this case, the measurement error model is defined in terms of  misclassification probabilities.
Consider a categorical variable $x$ with $K$ possible values, and denote the perturbed observed variable by $X$. The misclassification model can be parametrized through the misclassification probabilities for each category, defined  as:
\begin{equation}
p_{k'k}=P(X=k|x=k') \quad  k', k =1, \dots, K
\label{p_jk}
\end{equation}
So  $p_{k'k}$ represents the probability of observing category $k$ of $X$ for a record whose true latent category of  $x$ is $k'$.
The misclassification probabilities are collected into a $K \times K$ misclassification matrix $P$ whose diagonal elements $p_{kk}$ ($k=1,...,K)$ are the probabilities that no measurement error occurs for the $k$-th category of $x$.\\
In the context of classical measurement error in covariates, the regression model defined in terms of  the unobservable true covariate $x$ is supplemented by the misclassification model for $X$ given $x=k$, $k=1,...,K$, and then by a model for $x$:
\begin{equation}
\sum_{k', k} P(Y=y|x=k', \mathcal{B})P(X=k|x=k')P(x=k')
\label{carrol}
\end{equation}
where summation is done over all possible  combinations of levels of $(x,X)$ and $\mathcal{B}$ is the vector of the regression parameters. For each addend in (\ref{carrol}), the first component defines the underlying outcome model; the second one defines the error model for $X$ given the true covariates and the last term defines the distribution of the true covariate. This latter component is responsible for almost all the practical problems of implementation and model selection when the maximum likelihood method is employed.   
In the case of binary misclassified covariates, the maximum likelihood approach is relatively straightforward. For the general case ($K>2$), \citet{Kuchenhoff:2006} developed a  method, called MC-SIMEX, that corrects bias of model estimates when discrete covariates are misclassified.
Application to mixed effects models is discussed in \citet{Slate:2009}, who investigate the performance of MC-SIMEX  under a model with a discrete predictor measured with error and censored Gaussian responses. Their simulation study shows that even with the adjustment for bias allowed by the MC-SIMEX algorithm, considerable bias remains even with small misclassification probabilities. 
An important drawback of the method is that knowledge of the transition matrix $P$ is required. Excluding situations in which covariates are  perturbed on purpose for confidentiality reasons, this assumption is often unrealistic also in a small area context.

\section{Measurement error small area models}
\label{sec:02}
Models that are commonly used to derive small area estimators can be classified into two groups: area level models 
and unit-level models. Area level models relate the small area means to area-specific auxiliary variables. Such models are essential if unit-level data are not available. Unit-level models relate the unit values of the study variable to unit-specific auxiliary variables with known area means. In this paper we focus on unit-level models within a Bayesian framework \citep[see][for an up-to-date review]{Rao:2015}.\\
Suppose there are $m$ areas and let $N_{i}$ be the known population size of area $i$. We denote by $Y_{ij}$ the response of the $j-$th unit in the $i-$th area ($i=1,...,m$; $j=1,...,N_{i}$). A random sample of size $n_{i}$ is drawn from the $i-$th area. The goal is to predict the small area means $\Gamma_i=\frac{1}{N_i}\sum_{i=1}^{N_i} Y_{ij}$, $i=1,\dots, m$,  based on the available data.  To develop reliable estimates,  auxiliary information is introduced as covariates of a suitable regression model. The most common unit-level small area model is the so called nested error linear regression model
$$
Y_{ij}=\alpha + x_{ij}^\prime\beta+u_{i}+\epsilon_{ij} \qquad i=1,...,m; \;\;j=1,...,N_{i}
$$
The model is assumed to hold for the population as well as the sample units, e.g. under the hypothesis of no-selection bias.  \citet{Ghosh:2006} and \citet{Ghosh:2007} were the first to consider the problem of measurement error in small area models for unit-level data. Adopting a superpopulation approach to finite population sampling,  and assuming a single auxiliary variable defined at the area level, $x_i$, they model the response variable $Y$ as
\begin{equation}
\label{eq:1}
Y_{ij}=\alpha + \beta x_{i}+u_{i}+\epsilon_{ij} \hspace{1cm} i=1,...,m; \hspace{0.3cm}j=1,...,N_{i}.
\end{equation}
  $\epsilon_{ij}$ and $u_{i}$ are assumed independent,  $\epsilon_{ij} \iid N(0, \sigma^{2}_{e})$ and $u_{i} \iid N(0, \sigma^{2}_{u})$. 
To measure the true area-level covariate it is assumed that there are $R_{i}$ units in the $i-$th small area and that a random sample of size $r_{i}$ is taken from the $i-$th area, resulting in observable data $X_{il}$ ($l=1,..,r_i; i=1,...,m)$. For the sample, the measurement error model
\begin{equation}
\label{eq:2}
X_{il}=x_{i}+\eta_{il}, \qquad  \eta_{il}\iid N(0, \sigma^{2}_\eta)  \hspace{1cm} i=1,...,m; \hspace{0.3cm}l=1,...,r_{i}
\end{equation} 
is  assumed. Furthermore, $\epsilon_{ij}$, $u_{i}$ and $\eta_{il}$ are taken mutually independent. The model described in Equations (\ref{eq:1}) and (\ref{eq:2}) reduces to the one described in \citet{Ghosh:2006} when $R_{i}=N_{i}$ and $r_{i}=n_{i}$. 
 \citet{Ghosh:2006} also assumed that $x_{i} \iid N(\mu_{x},\sigma^{2}_x)$, defining the structural measurement error model.  
They considered both an empirical Bayes (EB) and a hierarchical Bayes (HB) approach to derive predictors of small area means. Under their empirical Bayes approach, \citet{Ghosh:2006} first derived a predictor for the   vector of $N_{i}-n_{i}$ units, conditional on the unknown parameters and the observed response, denoted as $Y^{(1)}$. In particular, for any unsampled $Y_{ij}$, $j=n_{i}+1,...,N_{i}$, they obtained
$$E[Y_{ij} | Y^{(1)},\beta,\alpha,\sigma^{2}_{e},\sigma^{2}_{u},\mu_{x},\sigma^{2}_{x},\sigma^{2}_{\eta}]=(1-B_{i})\bar{Y}_{i}+B_i(\alpha + \beta \mu_x)$$
where $B_{i}=\sigma^{2}_{e}/[\sigma^{2}_{e}+n_{i}(\sigma^{2}_{u}+\beta^2\sigma^{2}_{x})]$. The empirical Bayes predictor is obtained by replacing the unknown model parameters with their estimators.
\citet{Torabi:2009} extended the approach in \citet{Ghosh:2006} including sample information on the covariate values. 
 %
 \citet{Ghosh:2006} also proposed a fully Bayesian  approach; they define a hierarchical model based on equations (\ref{eq:1}) and (\ref{eq:2}),  specify vague prior distributions for all the model parameters,  and estimate posterior distributions via Gibbs sampling. \citet{Arima:2012} extended the above approach, proposing a Jeffreys' prior on the model parameters.

The aforementioned literature  considers the case in which the measurement error affects only continuous variables, according to the measurement error model of equations~(\ref{eq:1}) and (\ref{eq:2}). To allow for  auxiliary discrete covariates measured with error, we build on the model proposed in \citet{Ghosh:2006}. In particular, we model the  misclassification mechanism through an unknown transition matrix $P$ and estimate all the unknown parameters in a fully Bayesian framework. 
    
 \section{The proposed model}
\label{sec:03}
Consider a finite population, whose units are divided into $m$ small areas. As in the previous section, let 
$Y_{ij}$ be the value of the variable of interest associated
with the $j$-th unit 
in the  $i-$th area  and 
let the sample data be denoted by 
$y_{ij}$, $i=1,\dots, m$; $j=1,\dots ,n_{i}$. 
For each area, we consider the following covariates: $t_{ij}$ -- the vector of $p$ continuous or discrete covariates measured without error,  $x_{ij}$ -- the vector of  $h$ latent, misclassifed, discrete variables (with a total of $K$ categories), and finally $w_{i}$  -- the vector of $q$ latent continuous area-level covariates, measured with error. 

Denote by  $X_{ij}$ and $W_{ij}$ the observed vectors whose latent values are  $x_{ij}$ and $w_{i}$, respectively. We assume that the continuous covariates are perturbed independently and that misclassification only depends on the unobserved category of the latent variable, so if $h>1$ we assume independent misclassification. Without loss of generality, in what follows we assume $h=1$.
%

Following the notation in \citet{Ghosh:2006}, the proposed  measurement error model can be  
 written in the usual multi-stage way: for $ j=1,\dots,N_{i}$, $i=1,\dots, m$ and for $k,k'=1, \dots, K$
 \begin{align*}
\mbox{Stage 1. } &Y_{ij} = \theta_{ij} + e_{ij} &&  e_{ij} {\stackrel{\texttt{iid}}{\sim}} N(0,\sigma^{2}_{e})& \\
\mbox{Stage 2. } &\theta_{ij} = t_{ij}^{'}\delta + w_{i}^{'}\gamma + \sum_{k=1}^K I(x_{ij}=k)\beta_k+u_{i}  &&  u_{i} {\stackrel{\texttt{iid}}{\sim}} N(0,\sigma^{2}_{u}) &\\
\mbox{Stage 3. } & W_{ij}|w_{i}  {\stackrel{\texttt{iid}}{\sim}} N(w_{i},\Sigma_{\eta}) ,\; \Sigma_{\eta}=diag(\sigma^{2}_{\eta_1}, \dots, \sigma^{2}_{\eta_q}) &&  w_{i}  {\stackrel{\texttt{iid}}{\sim}} N(\mu_w,\Sigma_{w}) ,\; \Sigma_{w}=diag(\sigma^{2}_{w_{1}}, \dots, \sigma^{2}_{w_{q}})  &\\
& Pr(X_{ij}=k|x_{ij}=k')=p_{k'k}  &&p_{k'.}=(p_{k'1}, \dots, p_{k'K})\sim Dir(\alpha_{k',1}, \dots, \alpha_{k',K})\\
& Pr(x_{ij}=k')=\frac{1}{K} &&  &\\
 {\mbox{Stage 4. }} &   \beta,\delta,\gamma,\mu_w, \sigma^{2}_{e},\sigma^{2}_{u},\Sigma_{w},\Sigma_{\eta}
\text{ are, loosely speaking, a--priori}&&\!\!\!\!\!\! \text{}\text{ mutually independent.}&
\end{align*}
Stage 1 and 2 define the standard mixed effects model, expressed in terms of the unobservable covariates $w$ and $x$.  Note that  the intercept is not included in the model.
\\
Stage 3 defines the measurement error model for both continuous and discrete covariates:  as in \cite{Ghosh:2006}, we assume that each of the continuous observable covariates in $W_{ij}$ is modelled as a Gaussian variable centred at the true unobservable value $w_{il}$ with variability  $\sigma^{2}_{\eta_l}$, $l=1,\dots, q$. The vector $w_{i}$ is assumed normal with  mean  $\mu_{w}$ and variance-covariance matrix $\Sigma_w$, that, in line with the standard practice in regression models, may be assumed diagonal.
For the discrete covariates, the misclassification mechanism is specified according to the $K \times K$
matrix $P$, whose $(k',k)$ element, $p_{k'k}$, denotes the probability that the observable variable $X_{ij}$ takes the $k-$th category when the true unobservable variable $x_{ij}$ takes the $k'-$th category. We  assume that the misclassification probabilities are the same across subjects.
Row-wise, $P$ contains the conditional distributions of $X$ for the $K$ different values of $x$. We denote by $p_{k'.}$ the $k'-$th row of $P$, whose entries represent the probabilities $P(X_{ij}=k|x_{ij}=k')$, $k=1,\dots K$. Over each $p_{k'.}$, $k'=1,\dots,K$, we place a Dirichlet $Dir(\alpha_{k',1}, \dots, \alpha_{k',K})$  prior distribution, with known $\alpha_{k',1}, \dots, \alpha_{k',K}$.
We assume that all the categories of $x$ have the same prior probability $\frac{1}{K}$ to occur.  If prior information is available, this assumption can be easily 
relaxed considering different prior probabilities for the unobservable categories.
\\
In Stage 4  we assume $\beta~\sim~N(\mu_{\beta},\Sigma_{\beta}=diag(\sigma^2_\beta))$, $\delta~\sim~N(\mu_{\delta},\Sigma_{\delta}=diag(\sigma^2_{\delta}))$, $\gamma~\sim~N(\mu_{\gamma},\Sigma_{\gamma}=diag(\sigma^2_{\gamma}))$, 
 $\mu_{w} \sim N(\mu_{\mu_{w}},\Sigma_{\mu_{w}})$
$\sigma^{-2}_{u}~\sim~Gamma(a_{u},b_{u})$, $\sigma^{-2}_{e}~\sim~Gamma(a_{e},b_{e})$, $\sigma^{-2}_{\eta_l} \sim Gamma(a_{\eta_l},b_{\eta_l})$ and $\sigma^{-2}_{w_l} \sim Gamma(a_{w_l},b_{w_l})$, $l=1,\dots,q$.
Hyperparameters have been chosen to ensure flat priors (e.g. \cite{Ghosh:2006}). Finally, $(\alpha_{k',1}, \dots, \alpha_{k',K})$ are fixed as known.
%
%
According to the above assumptions, we can estimate the transition matrix $P$ jointly with all the other model parameters. 

\cite{Gelman:2006} points out that inverse Gamma prior\silvia{s} for the scale parameters in hierarchical models cannot be considered as non-informative.
In the context of small area  models, \cite{Ghosh:2006}, among the others, used inverse Gamma priors with parameter equal to 0.002 and state that the choice does not affect the estimates. Moreover, \cite{Arima:2012} discussed the use of non-informative priors in small area models and concluded that objective priors should be employed especially in small area framework where some objectivity is needed. They proposed an improper Jeffreys' prior that, under some mild conditions, leads to a proper prior. They also performed a simulation study and showed that when the number of observations increases ($n_{i}>3$), flat priors defined as in \cite{Ghosh:2006} and the proposed Jeffreys' prior perform very similarly  (see Scenario 3). As we will see in the next Sections, \silvia{the values of} $n_i$s in both the simulation study and the real data application allow us to use the prior in \cite{Ghosh:2006} as  non-informative.\\
Under the assumption that the model holds for the whole population as well as for the sample data $y_{ij}$,  e.g. under the hypothesis of no selection bias,  the specified model can be used to predict the small area means  $\Gamma_i=\frac{1}{N_i}\sum_{j=1}^{N_i}Y_{ij}$ given the available information; see e.g.  \citet{Rao:2015} (sec. 4.3 p.80, sec. 10.5 p. 362) and \citet{Datta:1998}.
More specifically, under the proposed model, estimates of the small area means can be derived from
\begin{equation}
\label{mu:i}
\mu_i=\frac{1}{N_i}\sum_{j=1}^{N_i}\theta_{ij}=\sum_{k=1}^K \beta_k F_{ik} + \gamma  w_{i}+\delta \bar t_{i}+u_i 
\end{equation}
where $F_{ik}={N_i}^{-1}\sum_{j=1}^{N_i} I(x_{ij}=k) $  are the relative frequencies of the $k-$th category of the variable $x$ for the $i-$th area in the population, and  $\bar t_{i}={N_i}^{-1}\sum_{j=1}^{N_i} t_{ij}$ is the vector of means of the auxiliary variable $t$ for the $i$-th area in the population. \\
\cite{Rao:2015} also noticed that prediction of $\Gamma_i$ does not exactly  correspond to predicting $\mu_i$, as in fact $\Gamma_i=\mu_i+\bar e_i$ with $\bar e_i={N_i}^{-1}\sum_{j=1}^{N_i} e_{ij}$, but when $N_i$ is large, the predictor of  the mixed effects  $\mu_i$ can be considered an appropriate predictor of $\Gamma_i$.\\
In measurement error problems, we usually do not observe the area-level population means for the covariates measured with error. As a consequence, $\mu_i$ in (\ref{mu:i}) involves quantities that  are likely not available when covariates are misclassified/mismeasured. However, \silvia{in analogy with}  \cite{Ghosh:2006} who replace the population mean  with the superpopulation mean of the covariate subject to error, 
we integrate the distribution of $\mu_i$ with respect to the posterior predictive distribution of $\Phi=(\beta,\delta, \gamma, u_i, x, w)$ given the sample data and  the population means of the auxiliary variables measured without error.  That is, we use the measurement error model to predict the distribution of the covariates $x$ and $w$.


Denote by $\Omega=(\beta, \delta, \gamma, \mu_{w},\sigma^{2}_{e}, \sigma^{2}_{u},\Sigma_{\eta},\Sigma_{w},P,x, w)$ the vector of the unknown parameters. The likelihood function is defined as
%
\begin{align}
\label{lik}
\nonumber
L&(\Omega; y) =\\
=& \sum_{k'=1}^{K} \frac{p_{k'k}}{K}  \prod_{i=1}^{m}\prod_{j=1}^{n_{i}}
\frac{1}{\sqrt{2\pi \sigma^{2}_{e}}} \exp \left(-\frac{1}{2\sigma^{2}_{e}}\left( y_{ij}-\sum_{k=1}^K I(x_{ij}=k)\beta_k -t_{ij}^{'}\delta - w_{i}^{'}\gamma - u_{i}  \right)^{2} \right)\nonumber\\
\nonumber
& \{2\pi |\Sigma_{\eta}|\}^{-\frac{n}{2}}\exp\left(-\frac{1}{2} \left(W_{ij}-w_{i}\right)'\Sigma_\eta^{-1} \left(W_{ij}-w_{i}\right) \right)\frac{1}{\sqrt{2\pi \sigma^{2}_{u}}}\exp\left(-\frac{1}{2\sigma^{2}_{u}} u_{i}^{2}\right)
\end{align}

According to the Bayes theorem, the posterior distribution of  $\Omega$ is proportional to the product of the likelihood 
above and the prior distributions specified in Stage 4. 
As the posterior distribution cannot be  derived analytically in closed form, we obtain samples from the posterior distribution using Gibbs sampling. Full conditional distributions to implement the Gibbs sampler are provided in the next subsection.
The MCMC output is exploited to estimate the elements of $\Omega$ as well as $\mu_i$, $i=1,\dots m$ in Equation (\ref{mu:i}).
%
%
%
\subsection{Computational details}
%
The Gibbs sampler requires drawing from the full conditional distributions of  {the elements of $\Omega$}
 given the remaining parameters and the data. The full conditional distributions are specified in the next equations.
%
Note that for the discrete variable(s) we adopt an Anova-type parametrization, denoting by $\mathbf x$ the  $(\sum_{i=1}^m n_i) \times K$ design matrix induced by the categorical variable(s); we recall that the intercept is not included in the model. Also, we denote by $\mathbf w$ the $(\sum_{i=1}^m n_i)\times q$ matrix of the latent continuous covariates subject to measurement error, obtained by stacking $n_i$ copies of the area-level covariate $w_i$, for each $i=1, \dots, m$, by $\mathbf T$ the $(\sum_{i=1}^m n_i)\times p$ matrix of the  continuous covariates measured without error, and finally  by $\mathbf Z$ the $( \sum_{i=1}^m  n_i) \times m$ known design matrix of zeros and ones needed to assign random effects to areas. 
\begin{itemize}
\item[(i)] $x_{ij}\vert \cdot \sim Multinomial (\pi_{1},...,\pi_{K})$ 
 where \\ 
$\pi_{k}=Pr(x_{ij}=k|\cdot) \propto p_{k,X_{ij}} \exp \left(-\frac{1}{2\sigma^{2}_{e}}  (y_{ij}-\beta_{k}-t_{ij}^{'}\delta - w_{i}^{'}\gamma-u_{i})^{2} \right)$
\item[(ii)] $\beta\vert \cdot \sim N\left( \left(\frac{{\mathbf x}'{\mathbf x}}{\sigma^{2}_{e}}+\frac{I}{\sigma^{2}_{\beta}} \right)^{-1}\left(\frac{{\mathbf x}' (y- \mathbf T \delta- \mathbf w \gamma  -\mathbf Z u)} {\sigma^{2}_{e}} +\frac{\mu_{\beta}}{\sigma^{2}_{\beta}}\right); \left(\frac{{\mathbf x}'{\mathbf x}}{\sigma^{2}_{e}}+\frac{I}{\sigma^{2}_{\beta}} \right)^{-1}\right)$
\item[(iii)] $\delta\vert \cdot \sim N\left( \left(\frac{\mathbf T'\mathbf T}{\sigma^{2}_{e}}+\frac{I}{\sigma^{2}_{\delta}} \right)^{-1}\left(\frac{\mathbf T' (y-{\mathbf x} \beta - \mathbf w \gamma  -\mathbf Zu)} {\sigma^{2}_{e}} +\frac{\mu_{\delta}}{\sigma^{2}_{\delta}}\right); \left(\frac{\mathbf T'\mathbf T}{\sigma^{2}_{e}}+\frac{I}{\sigma^{2}_{\delta}} \right)^{-1}\right)$
\item[(iv)] $\gamma\vert \cdot \sim N\left( \left(\frac{\mathbf w' \mathbf w}{\sigma^{2}_{e}}+\frac{I}{\sigma^{2}_{\gamma}} \right)^{-1}\left(\frac{\mathbf w' (y-{\mathbf x}\beta - \mathbf T\delta  -\mathbf Zu)} {\sigma^{2}_{e}} +\frac{\mu_{\gamma}}{\sigma^{2}_{\gamma}}\right); \left(\frac{\mathbf w'\mathbf w}{\sigma^{2}_{e}}+\frac{I}{\sigma^{2}_{\gamma}} \right)^{-1}\right)$
\item[(v)]$w_{i}\vert \cdot \sim N \left(\left(\frac{\gamma\gamma' }{\sigma^{2}_{e}}+{\Sigma_{w}^{-1}}+{\Sigma_{\eta}^{-1}}\right)^{-1} \!\!\left(
\gamma  \sum_{j=1}^{n_{i}}\frac{ s_{ij}}{\sigma^{2}_{e}}+
\Sigma^{-1}_{\eta} \sum_{j=1}^{n_{i}} W_{ij}+ {{\Sigma^{-1}_{w} \mu_{w}}} \right), \left(\frac{\gamma\gamma' }{\sigma^{2}_{e}}+{\Sigma^{-1}_{w}}+ \Sigma^{-1}_{\eta}\right)^{-1}  \right) $, 
 with $s_{ij}= y_{ij}-\sum_{k=1}^K I(x_{ij}=k)\beta_k- t_{ij}' \delta -u_i$
\item[(vi)] $\sigma^{-2}_{e}\vert \cdot \sim  Gamma \left(a_{e}+\frac{\sum_{i=1}^{m}n_{i}}{2}, b_{e}+\frac{\sum_{i=1}^{m}\sum_{j=1}^{n_{i}}(y_{ij}-\sum_{k=1}^K I(x_{ij}=k)\beta_k-t_{ij}^{'}\delta - w_{i}^{'}\gamma-u_{i})^{2}}{2}  \right)$
\item[(vii)] $\sigma^{-2}_{u}\vert \cdot \sim  Gamma \left(a_{u}+\frac{m}{2}, b_{u}+\frac{\sum_{i=1}^{n}u_{i}^{2}}{2}  \right)$

\item[(viii)] $\sigma^{-2}_{\eta_l}\vert \cdot \sim  Gamma\left ( a_{\eta_l}+\frac{\sum_{i=1}^m n_i}{2}, b_{\eta_l}+\frac{\sum_{i=1}^m \sum _{j=1}^{n_i} (W_{ijl}-w_{il})^2}{2}
 \right ), \; l=1,\dots, q$
\item[(ix)] $\mu_{w}\vert \cdot \sim  N((\Sigma_{w}^{-1}+\Sigma_{\mu_{w}}^{-1})^{-1}(\Sigma_{{w}}^{-1} \sum_{i=1}^m w_i+\Sigma_{\mu_{w}}^{-1}\mu_{\mu_{w}}),(\Sigma_{w}^{-1}+\Sigma_{\mu_{w}}^{-1})^{-1})$

\item[(x)]$\sigma^{-2}_{w_l}\vert \cdot \sim  Gamma\left ( a_{w_l}+\frac{\sum_{i=1}^m n_i}{2}, b_{w_l}+\frac{\sum_{i=1}^m  (w_{il}-\mu_{w_{l}})^2}{2}
 \right ), \; l=1,\dots, q$

\item[(xi)]$u_i\vert \cdot \sim N\left(\left(\frac{1}{\sigma^{2}_{u}} +\frac{n_i}{\sigma^{2}_{e}}\right)^{-1}\frac{\sum_{j=1}^{n_i}(y_{ij}-\sum_{k=1}^K I(x_{ij}=k)\beta_k-t_{ij}'\delta-w_{i}'\gamma)}{\sigma^{2}_{e}}, \left(\frac{1}{\sigma^{2}_{u}} +\frac{n_i}{\sigma^{2}_{e}}\right)^{-1}\right)$

\item[(xii)] $p_{k'.}=(p_{k'1}, \dots, p_{k'K})\vert \cdot\sim Dir(\alpha_{k',1}+\nu_{1,k'}, \dots, \alpha_{k',K}+\nu_{K,k'})$, $k'=1, \dots, K$, \\
where $\nu_{k,k'}$ is the number of occurrences such that $(X=k, x=k')$, for each $k=1, \dots, K$.
\end{itemize}

Model estimation has been implemented in the R environment \citep{R}. The code is available on request. 

 \section{Simulation study}
\label{sec:04}
%
The effect of measurement error in continuous covariates has been previously documented in \citet{Ghosh:2006} and \citet{Arima:2012}.  In order to emphasize the original contribution of the paper, 
here we focus on categorical covariates.
 
Following the simulation scheme in \cite{Ghosh:2006} and \cite{Torabi:2009}, we generated the following superpopulation:
\begin{equation}
\label{mod:sim}
y_{ij}=\theta_{ij} +e_{ij} = \sum_{k=1}^K I(x_{ij}=k)\beta_k+u_{i} +e_{ij}, \, j=1,\dots,N_{i}; \, i=1,\dots, m, 
\end{equation}
with $N_{i}$ ranging from $10^3$ to $10^6$. The number of areas was set to $m=20$.  We set  $K=3$ and sample  $x_{ij}$ iid from a uniform discrete distribution. We also set $\sigma^{2}_{e}=100$, $\sigma^{2}_{u}=16$ and  regression parameter $\beta=(50,5,-10)$.
\\
We generated 100 replicated samples from the model in Equation (\ref{mod:sim}). For each replicated sample, we select a random number of observations per area,  ranging from $3$ to $50$.
The actual observations $X$ in each sample were obtained by  perturbing the true categories of $x$ through a  transition matrix $P$ with diagonal entries $p_{kk}=p$ and off-diagonal entries all equal to $\frac{1-p_{kk}}{K-1}$, \silvia{$k=1, \dots, K$}. To investigate the effect of the misclassification, four levels of perturbation were set: $p=(0.5,0.6,0.7,0.8)$.  


Based on the simulated samples, we compare the proposed model, denoted by $M_{Prop}$, with the model $M_{True}$ that makes use of the true categories; in order to quantify the effect of ignoring the unknown misclassification mechanism, we also estimate the ``naive'' model $M_{Naive}$  that ignores the measurement error and uses the perturbed categories as if they were correct.
We focus on estimation of the model parameters $\beta, \sigma^{2}_{e},\sigma^{2}_u$ and on the model's ability at reconstructing the true categories of $x$. We also compare the predictions of the small area means  under each competing model.

The three models above share a subset of  hyperparameters, that at the estimation stage we fix at the same values, as follows:  $\mu_{\beta}=0$ and $\sigma^2_\beta=10^{4}$; also,  we set $a_{u}= b_{u}= a_{e}= b_{e}=0.001$; finally, to
ensure common prior means and variances for all the transition probabilities from a given category to any other category,
 for each $k=1, \dots K$, we chose symmetric Dirichlet priors  with  
 $\alpha_{k,h}=1/K, h=1,\dots, K$.  The aforementioned choice is the so called Perks' prior and is discussed in \citet{alvares} as  a non-informative prior when the quantity of interest is the whole vector of probabilities.  A sensitivity analysis (detailed in Section 1 of the Supplementary Materials) confirms the substantial robustness of the inferential conclusions with respect to the choice of  the hyperparameters.\\
 For each simulated data and for each model, we generate $10^{4}$ Markov chain Monte Carlo simulations, discarding the first half and then thinning the chains by taking one out of every 10 sampled values.
Table \ref{tab3} shows the assessment of the model's parameters in the simulation study. For each model and for each parameter, we report the posterior mean (Est), the relative bias (RB), the relative mean squared error (RMSE) and credible interval coverage (Cov) averaged over the 100 datasets. We remark that the elements of the vector $\beta$ should be assessed jointly, as they are associated to the indicator variables of the different categories of the same covariate.
\begin{table}[t]
\setlength{\tabcolsep}{2pt}
\begin{tabular}{ll@{\hspace*{1em}}cccc@{\hspace*{1em}}cccc@{\hspace*{1em}}cccc}
\multicolumn{2}{c}{} & \multicolumn{4}{c}{$M_{True}$} & \multicolumn{4}{c}{$M_{Prop}$} &\multicolumn{4}{c}{$M_{Naive}$}\\
\cmidrule(lr){3-6}\cmidrule(lr){7-10} \cmidrule(lr){11-14}
$p$  & & Est & RB & RMSE & Cov &Est & RB & RMSE & Cov & Est &RB & RMSE & Cov \\
\cmidrule{1-14}
$0.5$  & $\beta_1$ & 49.50 & -0.01 & 0.00 & 0.99 & 49.30 & -0.01 & 0.00 & 0.93 & 33.85 & -0.32 & 0.11 & 0.10 \\ 
  & $\beta_2$ & 4.31 & -0.08 & 0.10 & 0.97 & 2.49 & -1.50 & 1.49 & 0.65 & 3.03 & -0.39 & 0.45 & 0.82 \\ 
  & $\beta_3$ & -10.46 & 0.05 & 0.03 & 1.00 & -4.18 & -0.63 & 0.69 & 0.81 & 12.92 & -2.29 & 5.34 & 0.8 \\ 
  & $\sigma^{2}_{e}$ & 100.89 & 0.01 & 0.01 & 0.97 & 120.26 & 0.20 & 0.09 & 0.82 & 623.10 & 5.23 & 27.71 & 0.11 \\ 
  &$\sigma^{2}_{u}$ & 18.10 & 0.14 & 0.49 & 0.98 & 16.17 & 0.06 & 0.64 & 0.80 & 17.31 & 0.11 & 1.44 & 0.07 \\ 
\cmidrule[.1pt](lr){1-14}
$0.6$ &$\beta_{1}$ & 49.52 & 0.01 & 0.00 & 0.98 & 49.28 & -0.01 & 0.00 & 0.81 & 36.16 & -0.28 & 0.08 & 0.17 \\ 
  &  $\beta_{2}$ & 4.53 & 0.01 & 0.10 & 0.99 & 4.79 & -0.34 & 5.73 & 0.85 & 4.60 & -0.08 & 0.25 & 0.88 \\ 
   & $\beta_{3}$ & -10.46 & -0.05 & 0.02 & 0.99 & -7.94 & -0.21 & 0.12 & 0.81 & 8.28 & -1.83 & 3.42 & 0.27 \\ 
   & $\sigma^{2}_{e}$ & 100.07 & 0.00 & 0.01 & 0.96 & 119.38 & 0.19 & 0.07 & 0.85 & 578.66 & 4.79 & 23.18 & 0.28 \\ 
  &  $\sigma^{2}_{u}$ & 16.05 & 0.01 & 0.34 & 0.99 & 16.29 & 0.02 & 0.49 & 0.82 & 15.91 & -0.01 & 2.03 & 0.53 \\ 
 \cmidrule[.1pt](lr){1-14}
$0.7$ &$\beta_{1}$ & 48.93 & -0.02 & 0.00 & 0.96 & 48.72 & -0.03 & 0.00 & 0.95 & 36.99 & -0.26 & 0.07 & 0.18 \\ 
&   $\beta_{2}$& 3.89 & -0.02 & 0.12 & 0.97 & 4.49 & -0.24 & 0.23 & 0.82 & 6.46 & 0.29 & 0.29 & 0.62 \\ 
 &  $\beta_{3}$& -11.02 & 0.10 & 0.03 & 0.96 & -9.15 & -0.08 & 0.06 & 0.84 & 4.04 & -1.20 & 1.51 & 0.09 \\ 
   &  $\sigma^{2}_{e}$ & 102.72 & 0.03 & 0.01 & 0.94 & 119.76 & 0.20 & 0.08 & 0.83 & 517.62 & 4.18 & 17.76 & 0.17 \\ 
   & $\sigma^{2}_{u}$ & 15.10 & 0.06 & 0.35 & 0.95 & 14.25 & 0.06 & 0.41 & 0.94 & 13.66 & -0.05 & 1.20 & 0.54 \\ 
\cmidrule[.1pt](lr){1-14}
 $0.8$ & $\beta_{1}$  & 49.44 & -0.03 & 0.00 & 0.94 & 50.29 & -0.03 & 0.00 & 1.00 & 40.95 & -0.18 & 0.04 & 0.18 \\ 
   & $\beta_{2}$ & 6.44 & -0.12 & 0.10 & 0.98 & 3.36 & -0.37 & 0.22 & 0.92 & 7.03 & 0.47 & 0.38 & 0.84 \\ 
   &  $\beta_{3}$ & -9.79 & 0.06 & 0.05 & 0.96 & -9.29 & -0.01 & 0.01 & 0.90 & -2.98 & -0.70 & 0.58 & 0.23 \\ 
    & $\sigma^{2}_{e}$ & 101.41 & 0.01 & 0.01 & 0.91 & 112.31 & 0.12 & 0.03 & 0.85 & 432.04 & 3.32 & 11.24 & 0.09 \\ 
    &$\sigma^{2}_{u}$ & 16.57 & 0.05 & 0.36 & 0.99 & 17.46 & 0.02 & 0.29 & 0.99 & 14.25 & -0.04 & 1.23 & 0.84 \\   
  \end{tabular}
\caption{Simulation study: for each model parameter, the table reports the posterior mean (Est), the relative bias (RB), the relative mean squared error (RMSE) and credible interval coverage (Cov).  All the quantities are averaged over 100 datasets. In this simulation we set $K=3$. $\beta=(50,5,-10)$, $\sigma^{2}_{e}=100$, $\sigma^{2}_{u}=16$.}
\label{tab3}
\end{table}
%
 Table~\ref{tab3} allows us to compare our  proposal with the other approaches with respect to  estimation of model parameters.
 When the misclassification probability is relatively small ($p=0.8$ and $p=0.7$), the behaviour of the proposed model is very similar to the model involving the true covariates in terms of both bias and RMSE of the estimates. As expected, increasing the misclassification probability, the bias and variability of the { estimators} obtained under the proposed model tend to increase, but the coverage of the credible interval is still acceptable. On the other hand, ignoring the perturbation
has a significant effect on the estimates even when the probability of perturbation is small: increasing the perturbation, the variability  rapidly increases and so does the bias; by consequence, the credible intervals coverage is very low.
As stressed in \citet{Arima:2014}, measurement error in covariates introduces a sort of borrowing bias, which contrasts with the usual borrowing strength from the auxiliary information, typical of hierarchical models.

In some applications, it might be of interest to recover the true, unobserved, variable scores. For $p$ increasing from 0.5 to 0.8, the proportion of the true categories correctly inferred by the proposed model is equal to  $0.711, 0.778, 0.829$ and $0.892$: despite the use of a symmetric Dirichlet prior on the transition probabilities $p_{k'.}$, the information conveyed by the model allows us to correctly reconstruct a proportion of the $x_{ij}$s that is even larger than prescribed by the perturbation scheme.

\begin{figure}
{\centering
\hspace*{-2em}\includegraphics[width=0.58\textwidth]{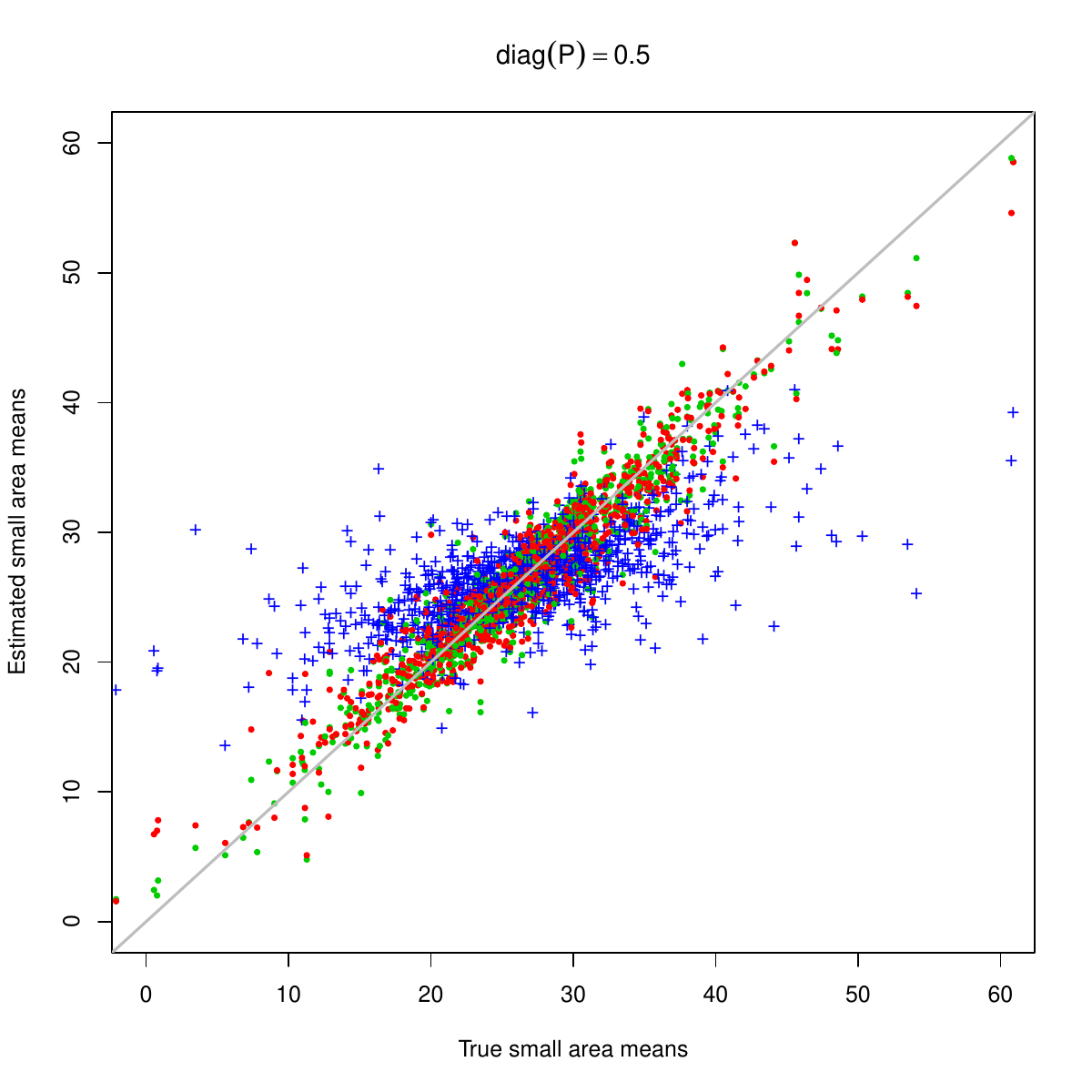}\hspace*{-1em}
\includegraphics[width=0.58\textwidth]{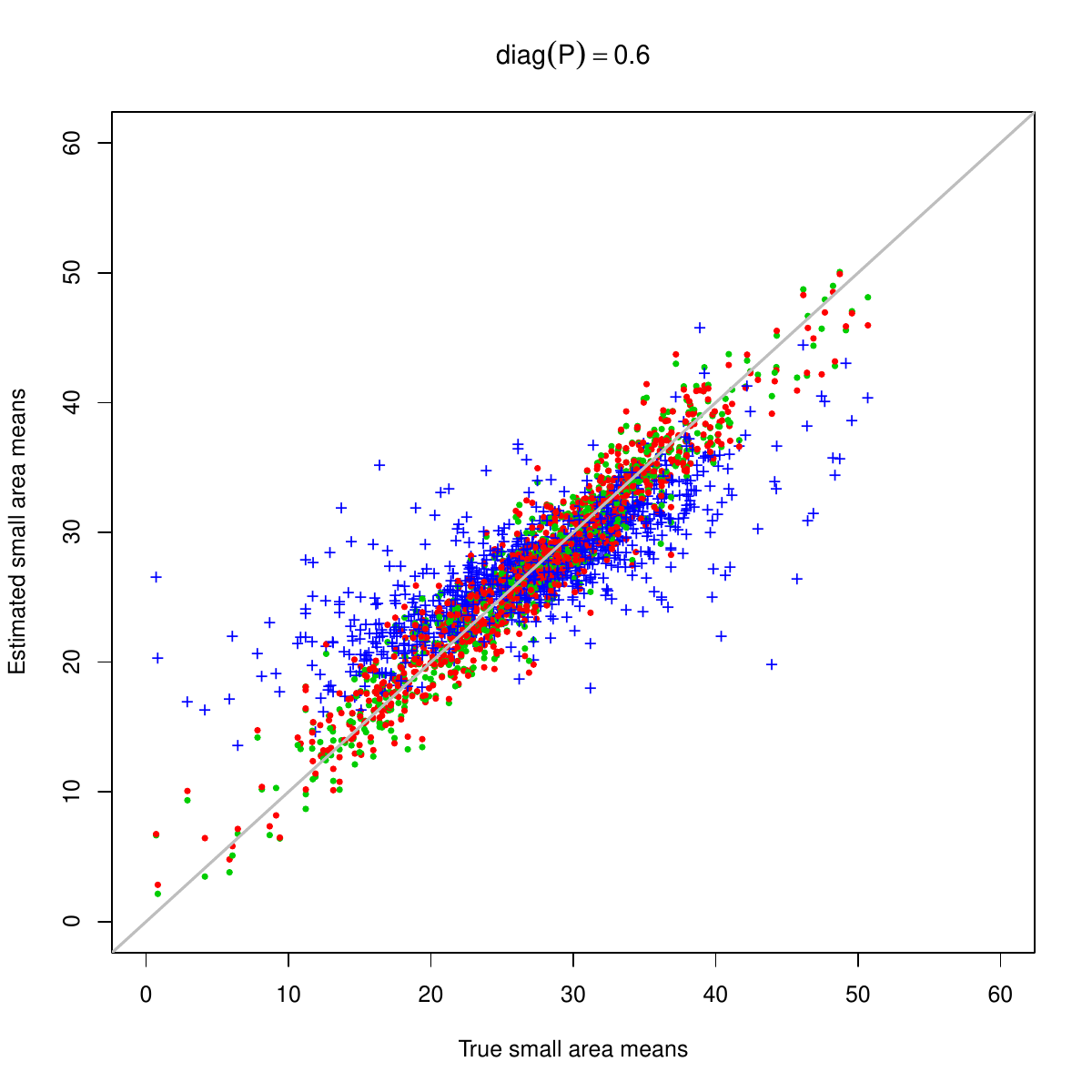}\\
\hspace*{-2em}\includegraphics[width=0.58\textwidth]{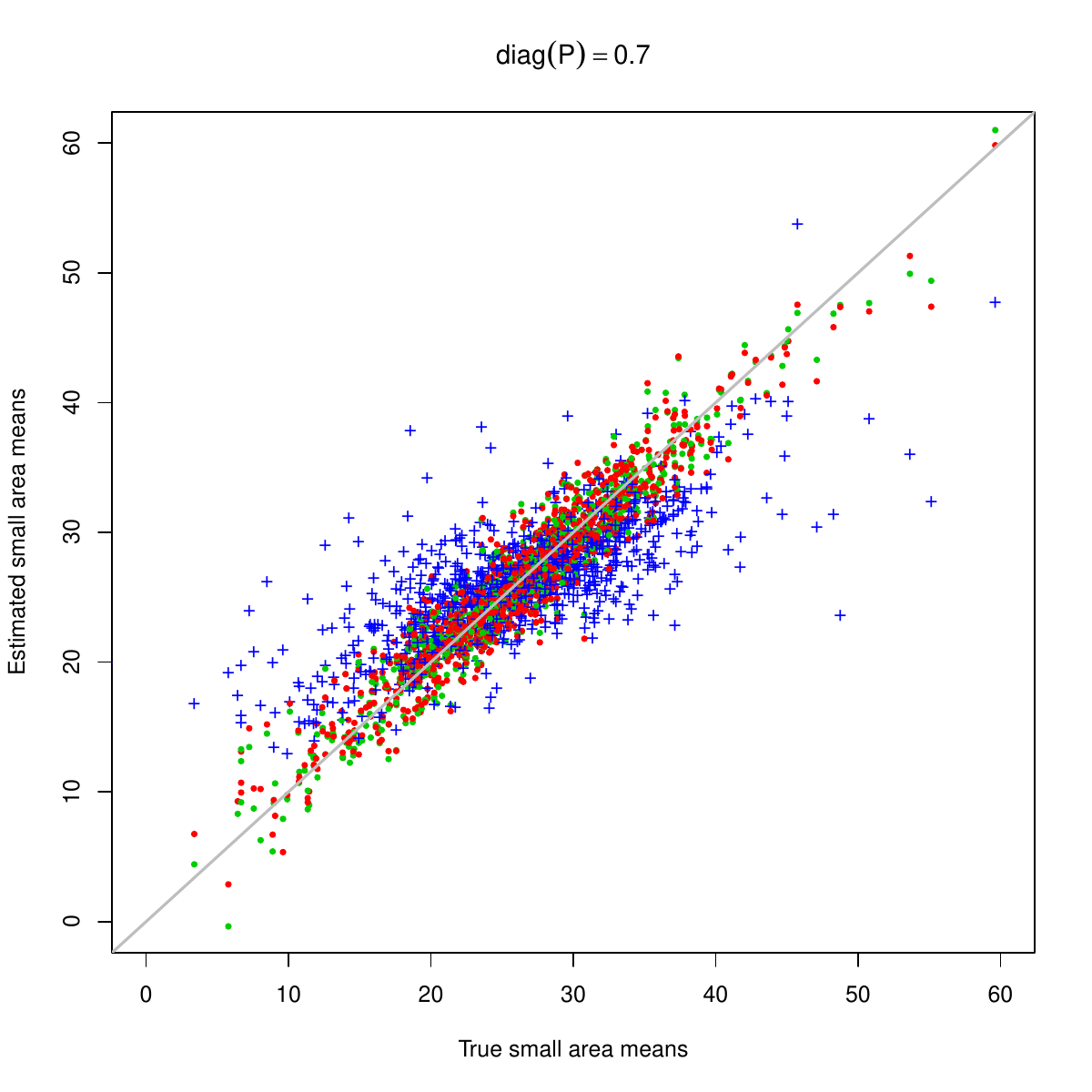}\hspace*{-1em}
\includegraphics[width=0.58\textwidth]{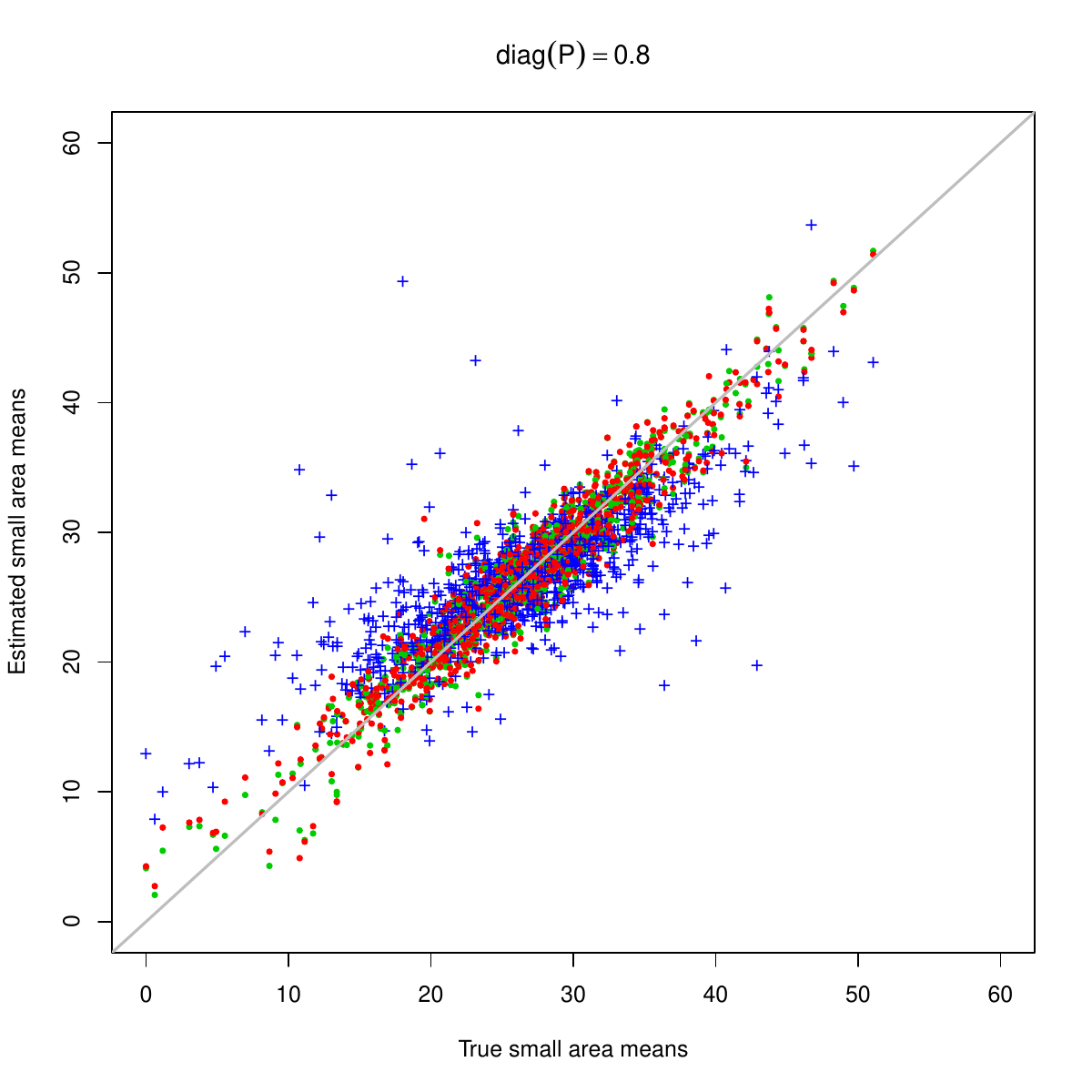}
}
\caption{Simulation study: estimated small area means  versus the true small area means under different misclassification scenarios ($p=0.5,0.6,0.7,0.8$). We highlighted with red points the small area means obtained with the proposed model ($M_{Prop})$, with green points and blue crosses those obtained respectively with the true model ($M_{True}$) and the model ignoring the measurement error ($M_{Naive}$).}\label{fignew:04}
\end{figure}

\begin{figure}

{\centering
\hspace*{-2em}\includegraphics[width=0.58\textwidth]{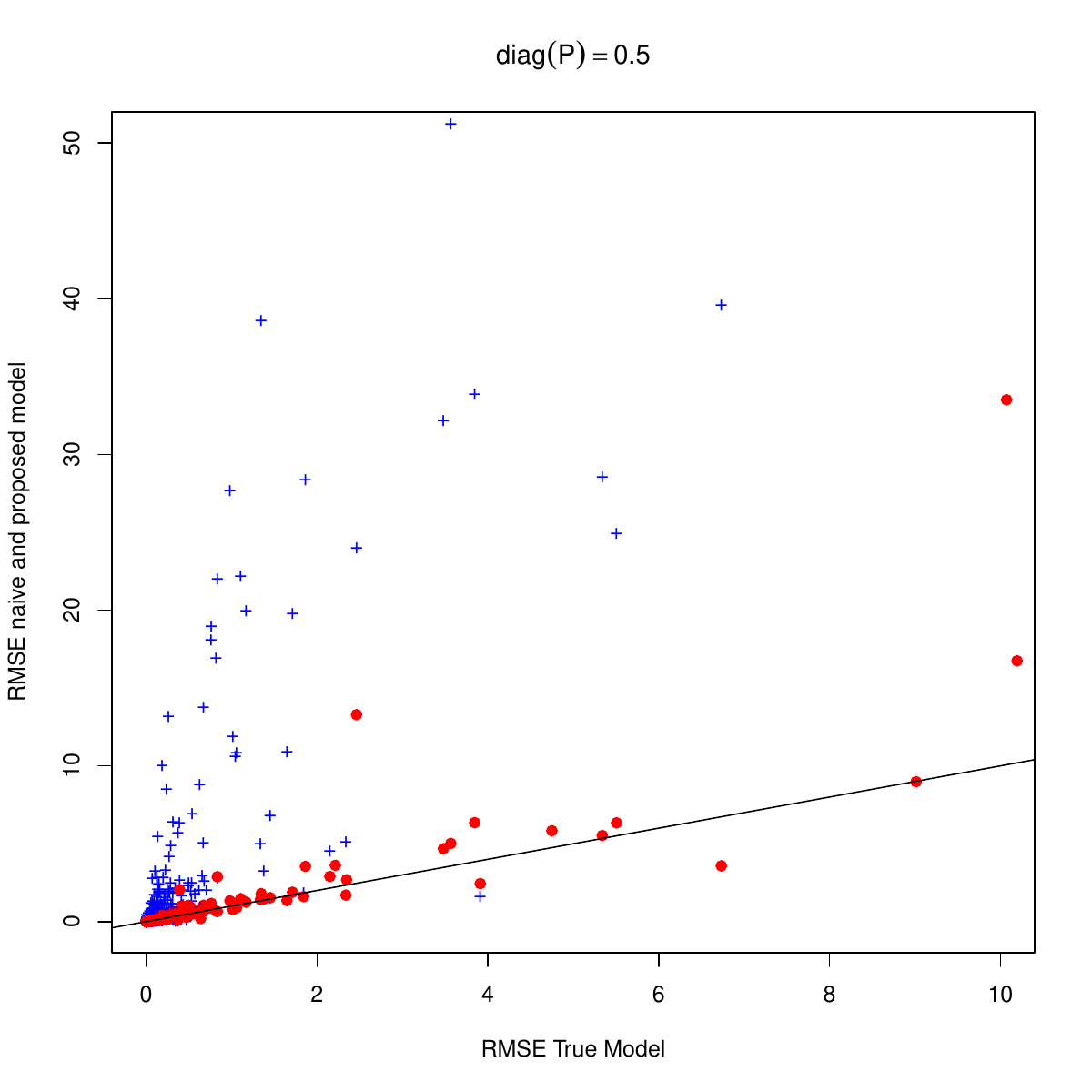}\hspace*{-1em}
\includegraphics[width=0.58\textwidth]{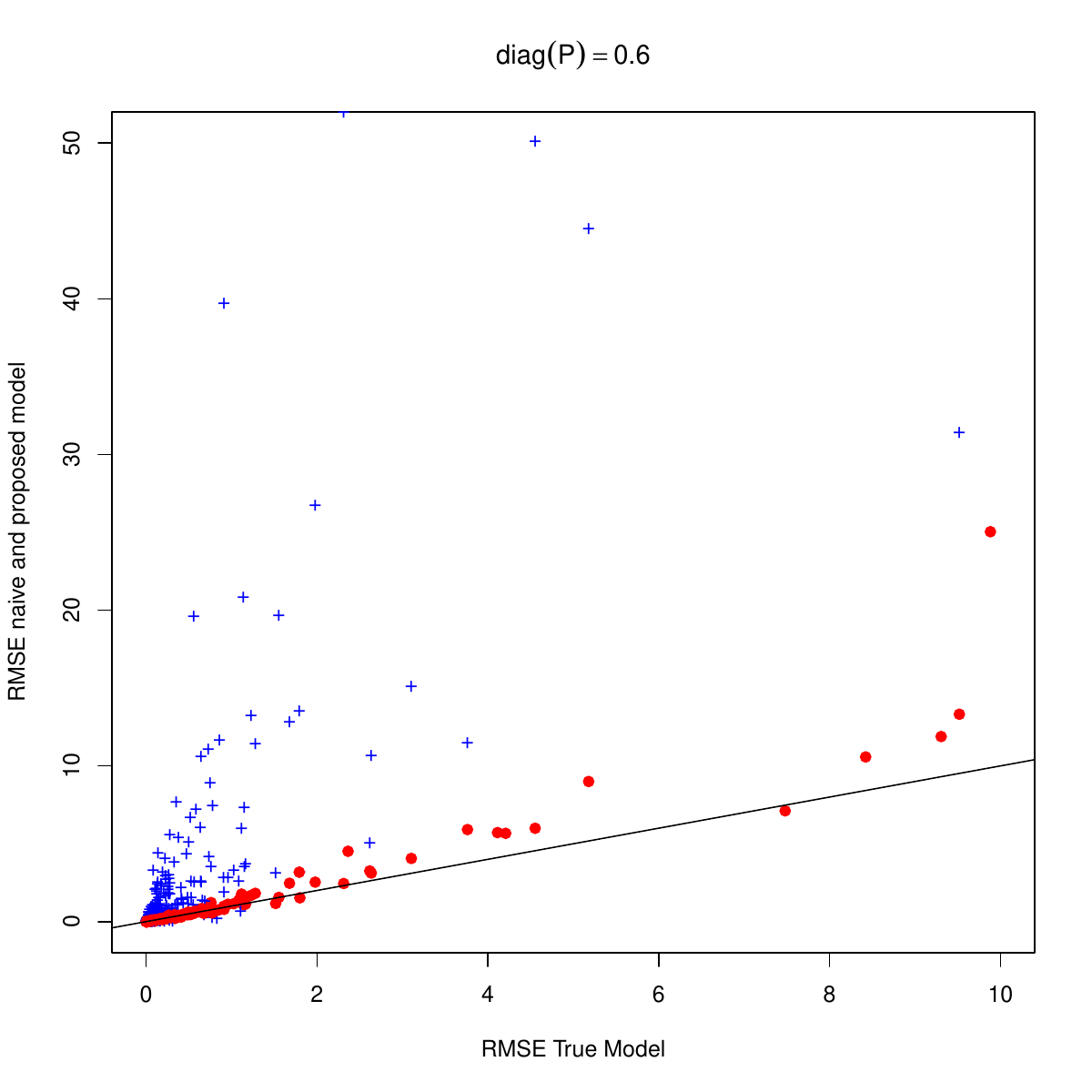}\\
\hspace*{-2em}\includegraphics[width=0.58\textwidth]{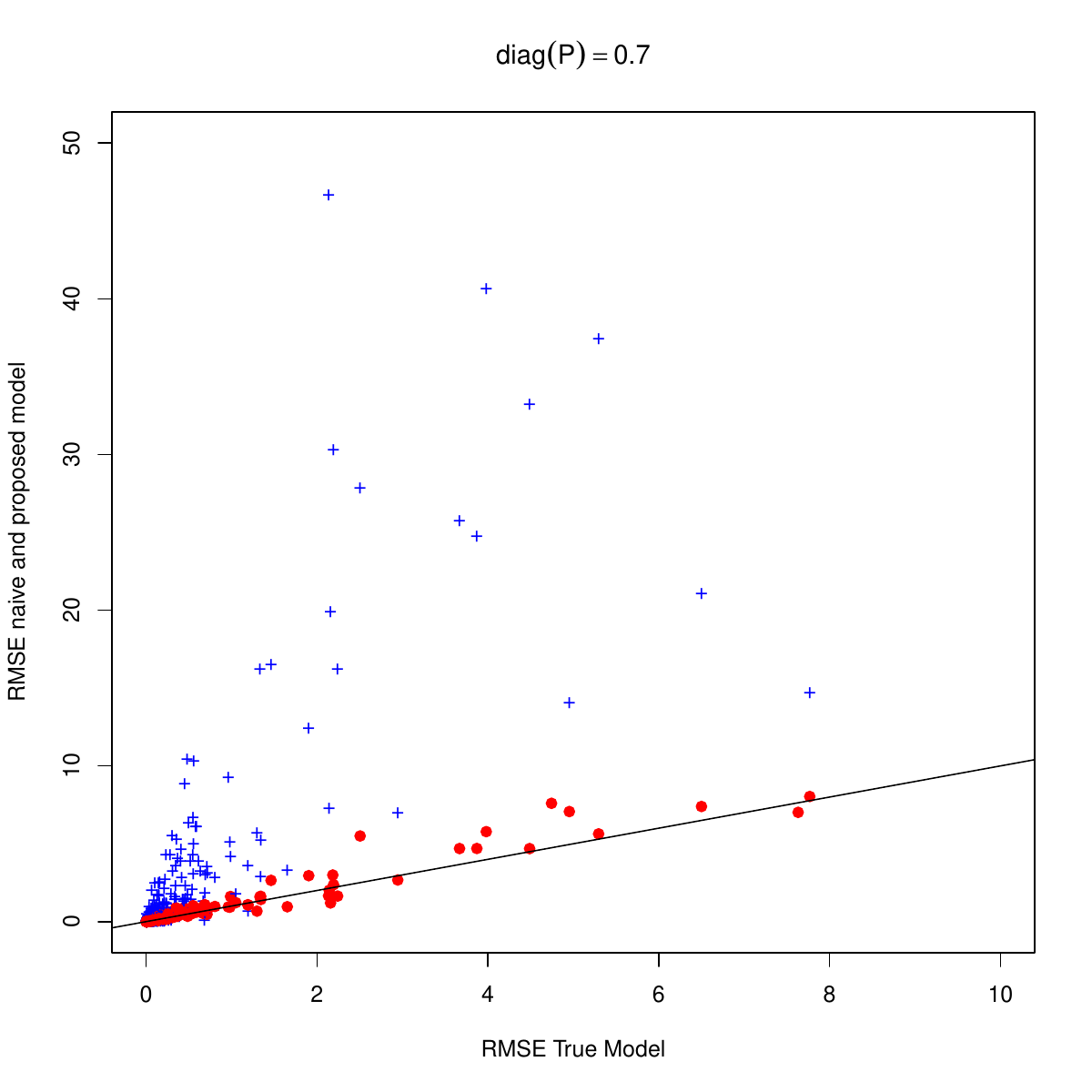}\hspace*{-1em}
\includegraphics[width=0.58\textwidth]{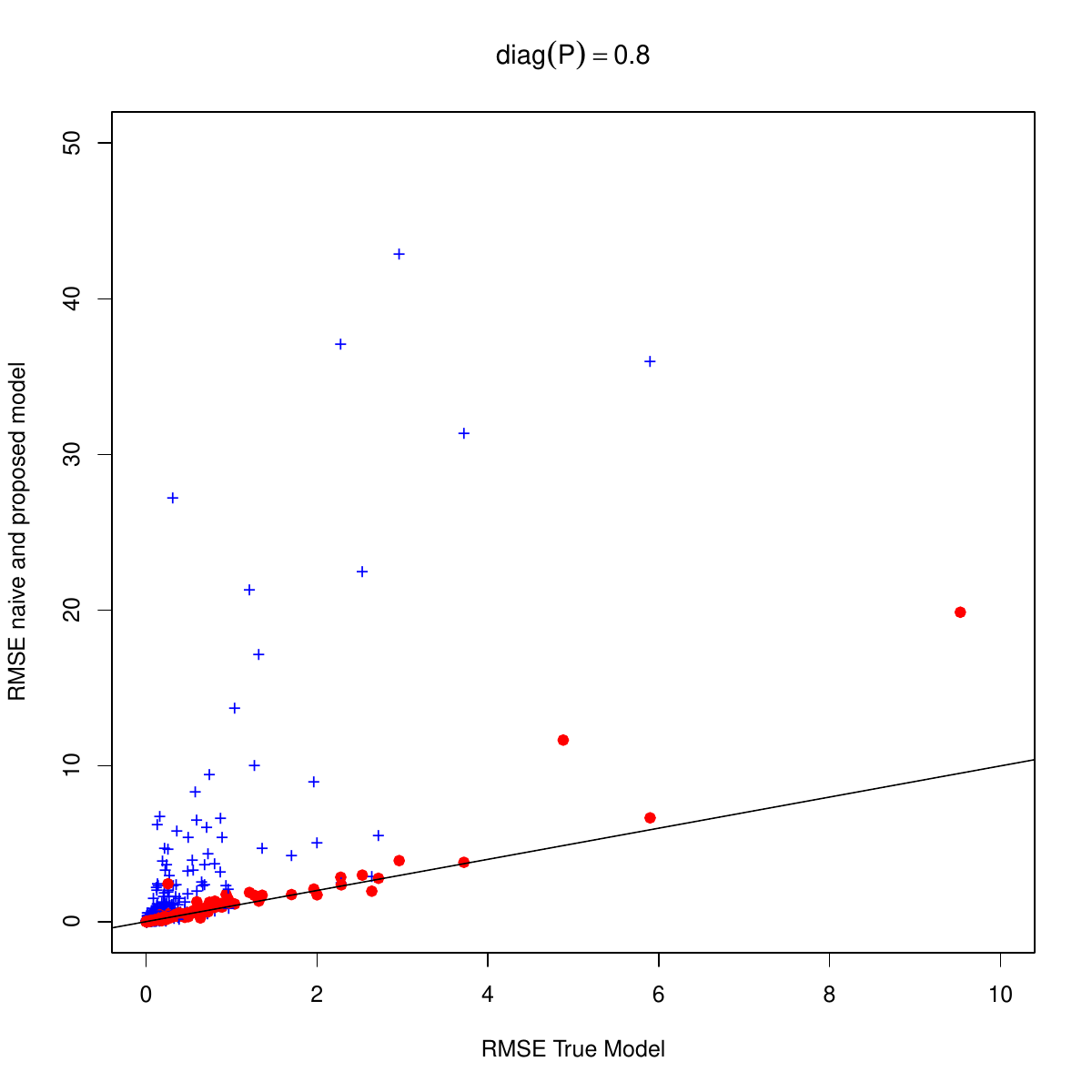}
}
\caption{Simulation study: RMSE (multiplied by 100) of the small area means under different misclassification scenarios ($p=0.5,0.6,0.7,0.8$).We plot the RMSE of the small area means obtained under the proposed model ($M_{Prop}$, red points) and under the model ignoring the measurement error ($M_{Naive}$, blue crosses) versus the RMSE of the small area means estimated with the true model ($M_{True}$).}\label{fignew:05}
\end{figure}

Figure \ref{fignew:04}  shows the estimated small area means with the true model $M_{True}$, the proposed model $M_{Prop}$ and with the model ignoring the measurement error  $M_{Naive}$ versus the true small area means. Figure \ref{fignew:05} displays the RMSE of the small area mean estimates obtained under $M_{Prop}$ and under $M_{Naive}$ plotted against the RMSE of the small area mean estimates obtained under $M_{True}$. The RMSEs computed according to the true small area means are always smaller under the measurement error model,  which in fact is the data generating model, than under the model that ignores the measurement error. These findings reflect a bias-variance trade-off which favours  the proposed model for the perturbation levels at hand. 
Also, the RMSEs are very similar to those of the model involving the true covariates when the misclassification probability is low.   When $p$ decreases (higher misclassification), the small area prediction errors increase, but they are much smaller than those produced by the model ignoring the measurement error. 
The credible interval coverage (shown in Table 3 of the Supplementary Materials) of the proposed model is close to the nominal level (95\%) even for large misclassification errors, whereas the naive model always shows a very poor performance. Further investigations (not shown here) reveal that the bias reduction balances the variance inflation up to very mild perturbation levels ($p=0.95$; the extreme case of no misclassification, e.g. $p=1$, is discussed below).  

\begin{figure}
{\centering
\hspace*{-2em}\includegraphics[width=0.58\textwidth]{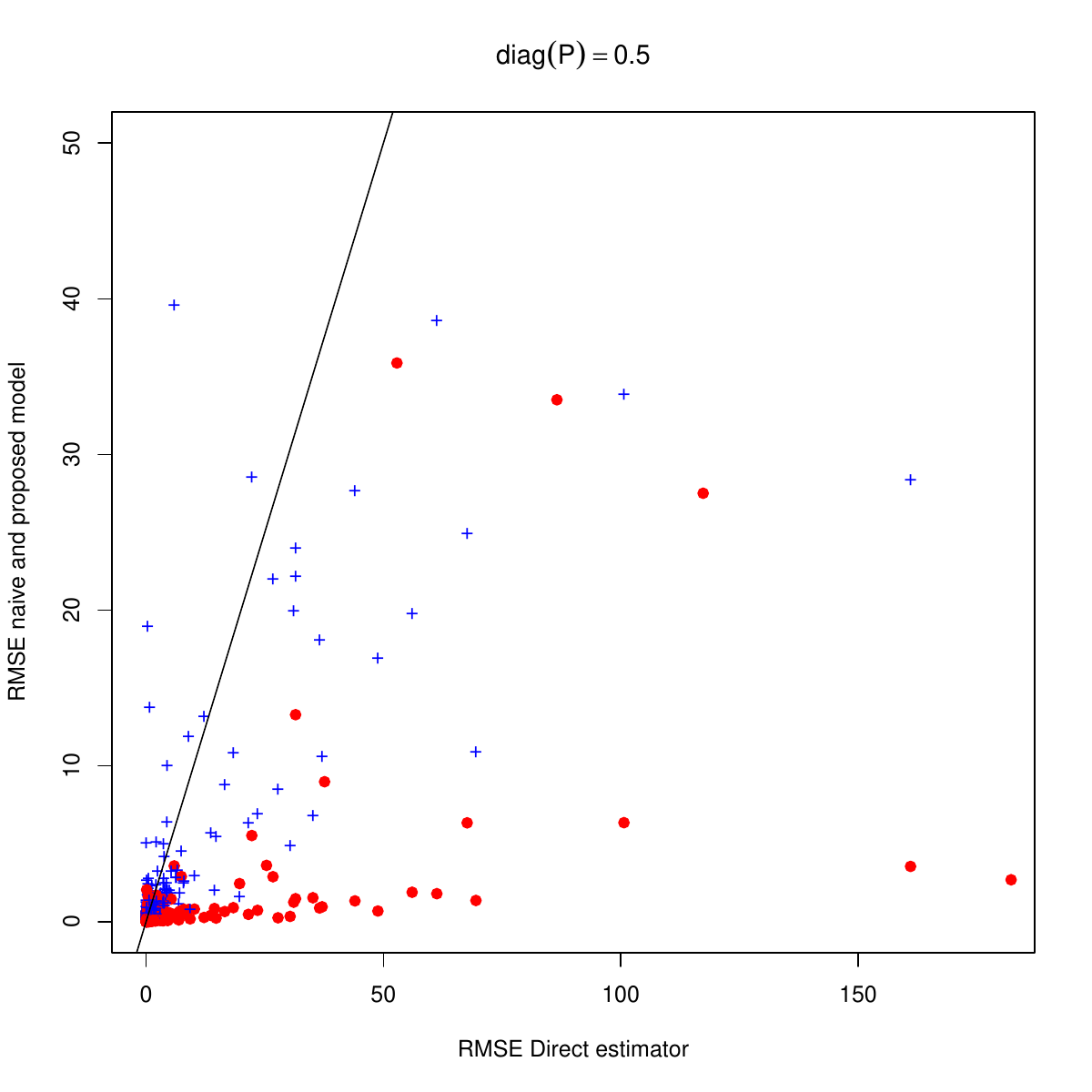}\hspace*{-1em}
\includegraphics[width=0.58\textwidth]{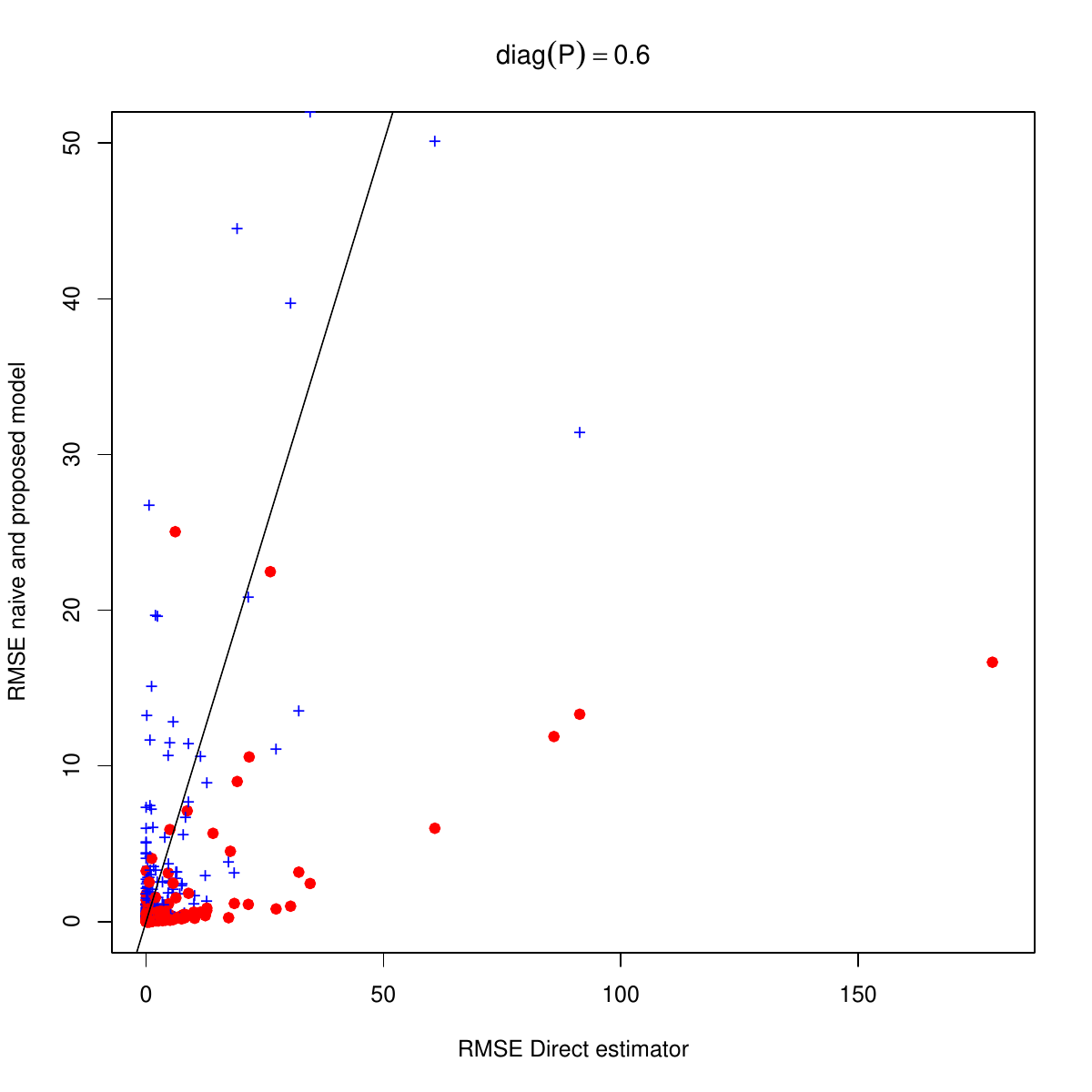}\\
\hspace*{-2em}\includegraphics[width=0.58\textwidth]{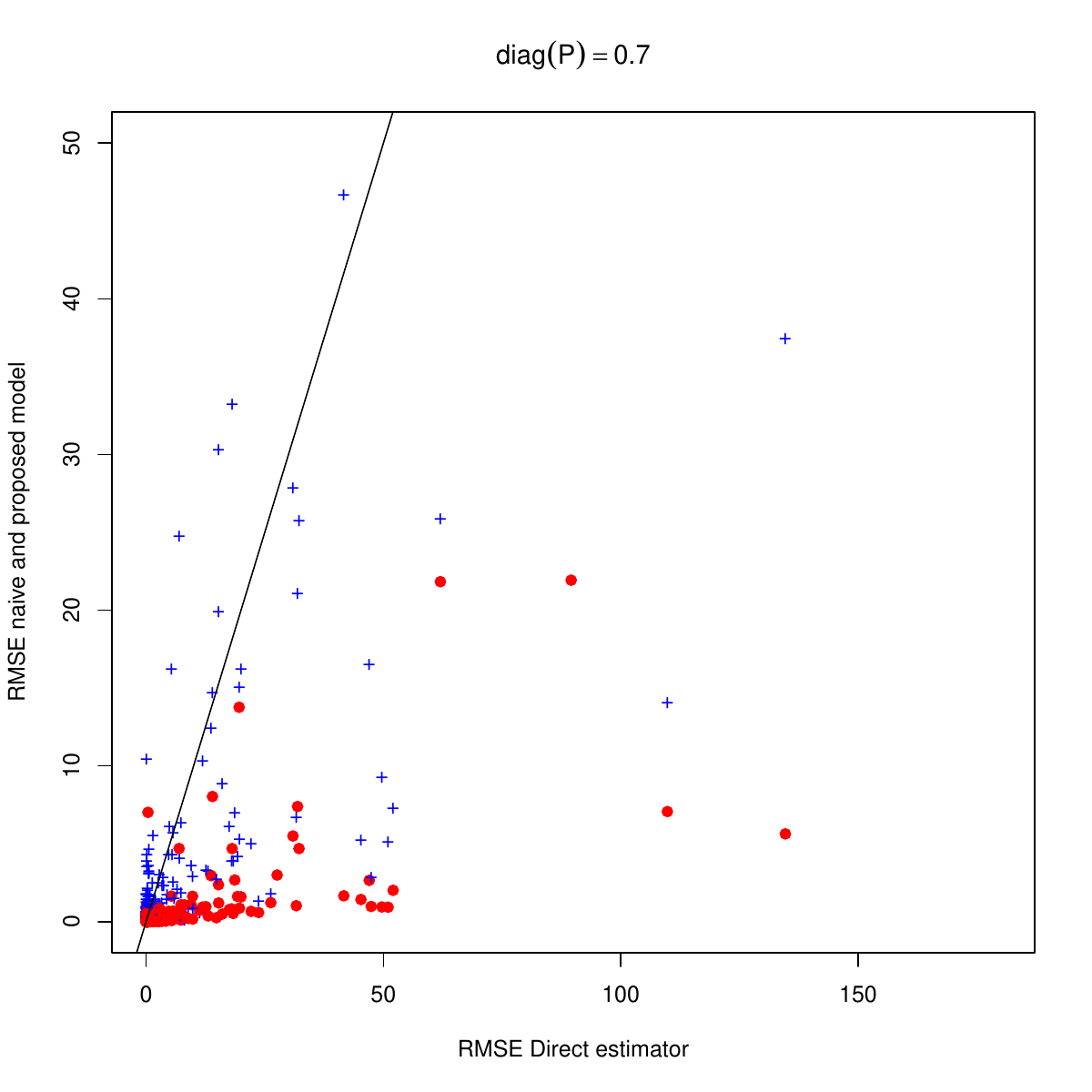}\hspace*{-1em}
\includegraphics[width=0.58\textwidth]{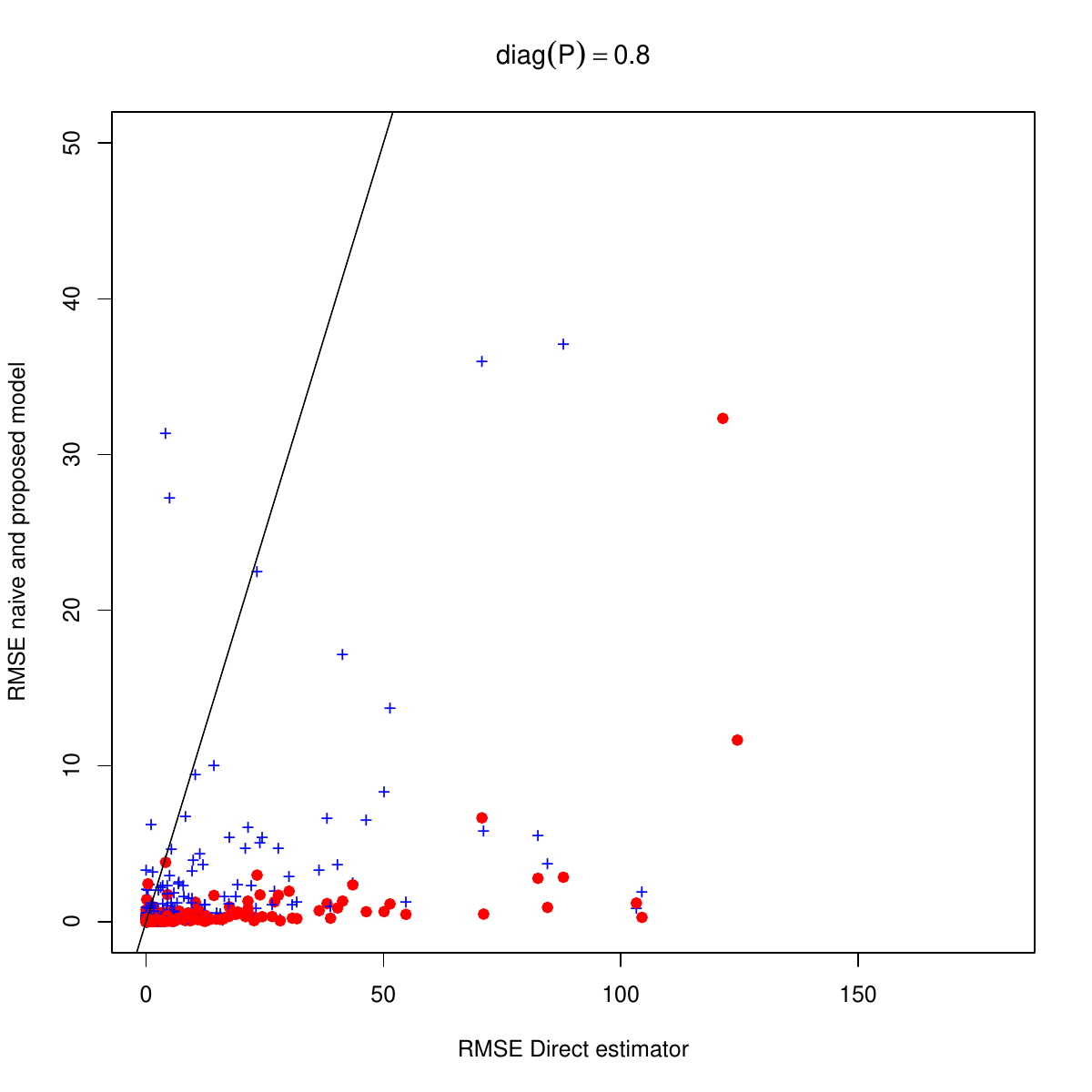}
}
\caption{Simulation study: RMSE (multiplied by 100) of the small area means under different misclassification scenarios ($p=0.5,0.6,0.7,0.8$). We plot the RMSE of the small area means obtained with the proposed model  ($M_{Prop}$, red points)  and with the model ignoring the measurement error ($M_{Naive}$, blue crosses) versus the RMSE of the direct estimator.} \label{fignew:06}
\end{figure}

\begin{figure}[ht]
\centering \hspace*{-2em}\includegraphics[width=0.58\textwidth]{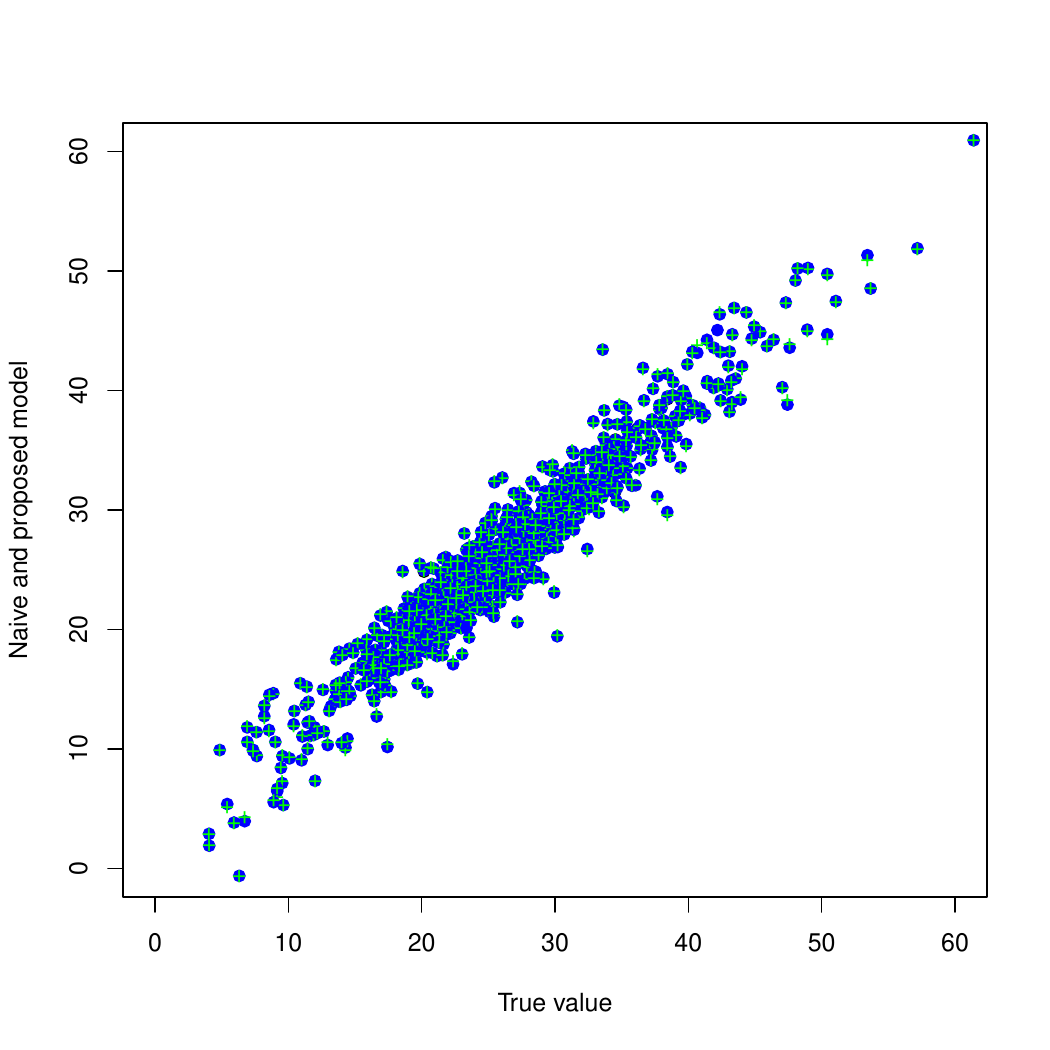}\hspace*{-1em}
\includegraphics[width=0.58\textwidth]{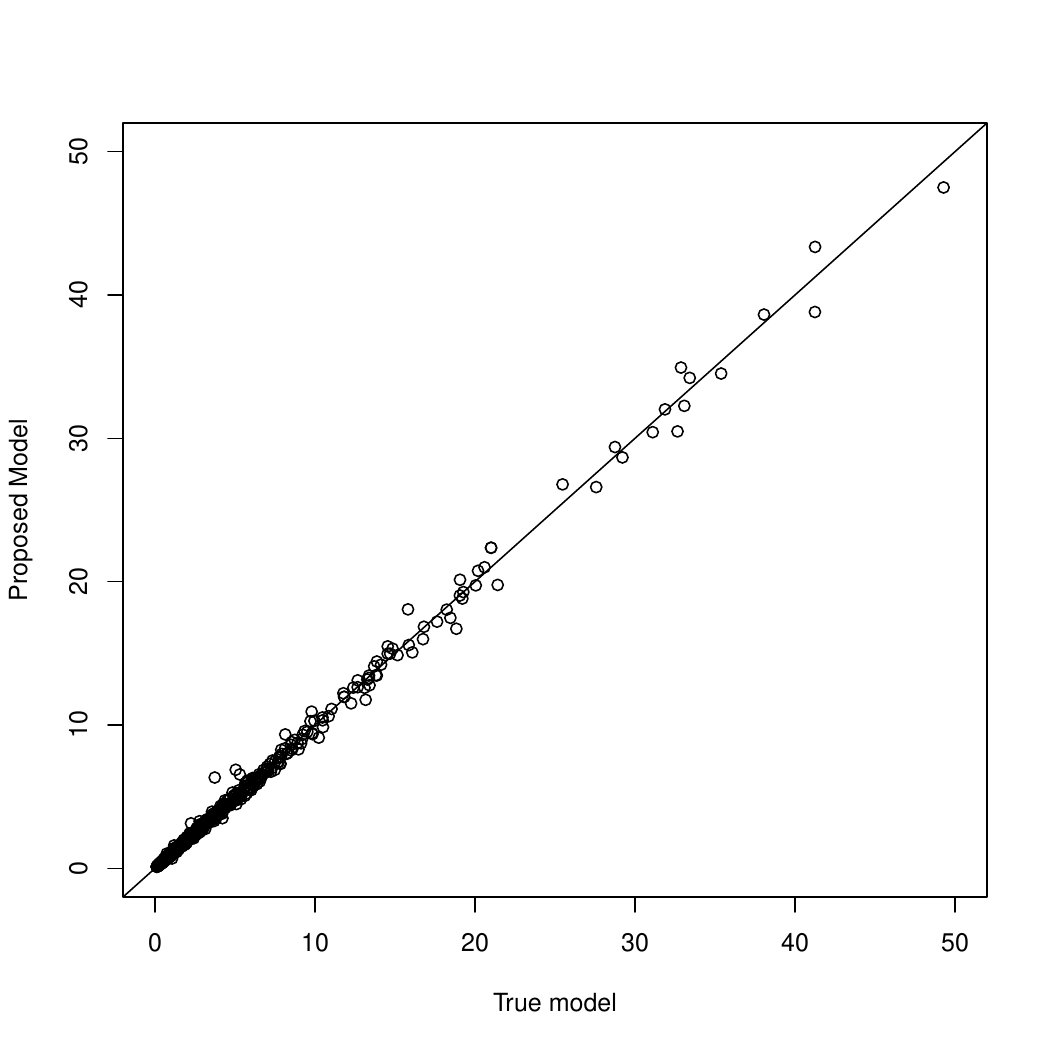}
\caption{Simulation study: no misclassification scenario. Left panel: small area means estimated with the true model (green crosses) and the proposed model (blue points) versus the true small area means. Right panel: RMSE of the small area means estimated with the proposed versus the true model.}\label{fig:07}
\end{figure}

Figure \ref{fignew:06} compares the RMSE of the small area means obtained under the proposed model and under the model ignoring the measurement error with the RMSE of the direct estimator.
Compared to the model based predictors, the simulation study shows that the direct estimates have, as expected, larger variability and tend to be in disagreement with the true values (large RMSEs). Moreover, the posterior standard deviations of the small area means are larger under the measurement error model than under the standard one, with increasing variability for higher perturbation levels. This may be expected, as an extra source of variability is introduced in $M_{Prop}$ (see Table 1 of the Supplementary Materials).
\begin{table}[tb]
\centering
\setlength{\tabcolsep}{2pt}
\begin{tabular}{ll@{\hspace*{1em}}cccc@{\hspace*{1em}}cccc}
\multicolumn{2}{c}{} & \multicolumn{4}{c}{$M_{True}$} & \multicolumn{4}{c}{$M_{Prop}$}\\
\cmidrule(lr){3-6}\cmidrule(lr){7-10}
$p$  & & Est & RB & RMSE & Cov &Est & RB & RMSE & Cov \\
$1.0$ & $\beta_{1}$ & 49.58 & -0.01 & 0.00 & 1.00 & 48.78 & -0.02 & 0.00 & 0.92 \\
& $\beta_{2}$ & 4.47 & -0.11 & 0.09 & 1.00 & 4.82 & -0.24 & 0.14 & 0.84 \\
& $\beta_{3}$ & -10.46 & 0.05 & 0.03 & 0.98 & -11.21 & 0.12 & 0.03 & 0.94\\
& $\sigma^{2}_{e}$& 100.86 & 0.01 & 0.01 & 0.98 & 100.70 & 0.01 & 0.03 & 0.98 \\
& $\sigma^{2}_{u}$ & 14.88 & -0.07 & 0.30 & 0.99 & 14.27 & -0.28 & 0.27 & 0.92 \\
\hline
\end{tabular}
\caption{Simulation study: with the same settings as in Table~\ref{tab3}, the table reports the posterior mean (Est), the relative bias (RB), the relative mean squared error (RMSE) and credible interval coverage (Cov) when there is no perturbation ($p=1.0$) for the true and the proposed model. }
\label{tab-NoErr}
\end{table}
{\blue To follow up on analysing the performance of the proposed model when the perturbation level decreases, we also simulate data under  the extreme scenario of } 
no misclassification, i.e. the same settings as in the previous simulation scheme, but specifying $p=1$ and, as a consequence, $X\equiv x$. 
Table~\ref{tab-NoErr} compares the parameter estimates obtained under the true and the proposed model, respectively (notice that in this case the naive and the true model are the same). As expected, the proposed model satisfactorily estimates the model parameters. At the same time the impact on the relative bias and mean squared error is in line with the trend (decreasing with $p$) already shown in Table~\ref{tab3}. Also, the matrix $P$ is coherently estimated  and the proportion of true categories correctly inferred is equal to 98.4\%.  {Finally,  Figure \ref{fig:07} shows small area means predictions (left panel) and the RMSE of the small area mean predictors (right panel) estimated with the proposed model and the true model, when the data generating model has no misclassification in the auxiliary variable. As expected, small area predictions are coherently estimated with the proposed model and no substantial differences may be grasped in terms of RMSE.}
As regards the error in covariates, we conclude that the proposed approach is robust against model misspecification.

\section{Ethiopia DHS data application}
\label{sec:05}

We consider data from the 2011 Ethiopia DHS, 
 a nationally representative survey of 16,515 women aged 15-49 and 14,110 men aged 15-59.  Data are available at \verb[www.measuredhs.com[. 
The Ethiopia DHS sample is a two-stage stratified cluster sample. It is designed to produce representative estimates of key indicators for the country as a whole, for the urban and the rural areas separately, and for each of the eleven regions of the Country (nine regional states, namely Tigray, Affar, Amhara, Oromiya,  Somali,  Benishangul-Gumuz, SNNP, Gambela, Harari, and
two city administrations, Addis Ababa and Dire-Dawa); the latter represent the 11 domains of interest in our application.
The first stage primary sampling units (EA, census enumeration areas defined for 2007 Census) were stratified by region and urban/rural characteristics. 
In light of the variability across regions of the household distribution, non-proportional allocation of the sample to the different regions and to their urban and rural areas is preferred. At the second stage, a fixed number of households has been randomly sampled from the primary sampling units. See the Country-specific DHS documentation~\citep{EDHS} for further details on the sampling design.

Complete survey data were available for a total of $\sum_i n_i = 15515$ women,  with $n_i$ ranging from 813 (Somali)  to 2066 (Oromiya). Such sample sizes, although not typical of small area applications, should be considered in light of the total population (over 96 millions of inhabitants in 2014, which makes Ethiopia the second most populated country in Sub-Saharan Africa) and of the strong geographical variability; moreover these regions represent unplanned domains for female population for the variable of interest. For the above reasons, the problem might be framed within a small area estimation context. 
We considered women's body mass index (BMI) (kg/m2)  as a measure of their nutritional status. 
%
Several studies \citep[see e.g.][]{DHS,DHS3} have investigated  the role of age, marital status, religion, occupation, education attainment and living standard as potential determinants of  women's nutritional status in Ethiopia. They also highlighted strong regional disparities and relevant  differences between urban and rural areas. The latter often represent the poorest areas, and are 
characterized by high incidence of infectious diseases, 
 low access to improved water sources,  high exposure to  natural hazards, scattered health service provisions and, finally,  lack of access to education and early marriage for women. 
The living condition was measured through the household wealth index quintile, a composite measure of a household's cumulative living standard that is available from the DHS data.
Following the previous literature, we consider a unit-level small area model with area-specific random effects that capture the possible regional differences in BMI levels. This allows us to estimate BMI levels at the area level, as well as to assess the covariates' effects on BMI. The model exploits the relationship between women's BMI
 and the following covariates: area of residence (urban vs rural centres), wealth index, educational attainment (with three categories: primary, secondary, higher education), number of children ever born (parity) and age.  
We notice that  the wealth index  is built on the survey data through a complex procedure which makes it a variable subject to several sources of error. 
The wealth index is \silvia{indeed} built from information on  asset ownership,
housing characteristics and water and sanitation facilities; it is obtained via a three-step procedure, based on principal components analysis, designed to take better account of urban-rural differences in scores and indicators of wealth \citep{wealth:ind,EDHS}.  
National-level wealth quintiles (from lowest to highest) are obtained by assigning the household score to each {\em de jure} household member, ranking each person in the population by his or her score, and then dividing the ranking into five equal categories, each comprising 20 percent of the population. By consequence, we treat the wealth index quintile as a categorical covariate subject to misclassification.
Among regions, the wealth quintile distribution varies greatly. A relatively high percentage of the population in the most urbanized regions is in the highest wealth quintile, while a significant proportion of the population in the
more rural regions is in the lowest  quintile, as in Affar (57\%), Somali (44\%), and
Gambela (35\%). 

We also consider age, being self reported, as  a continuous variable observed with error.

To accommodate error in variables 
we apply  the methodology described in Section~\ref{sec:03}. Our aim is to investigate the effect of neglecting  measurement error in both categorical and continuous covariates on the assessment of the regression effects and on the  prediction of area-specific BMI mean levels. 
In order to assess the effect of measurement error, we estimate  the proposed model, along with  its no-measurement-error counterpart.  We assume that all the regression effects {\blue a priori have} zero mean and variances equal to $10^4$ and set $a_{u}, b_{u}, a_{e}, b_{e}, a_{\eta}, b_{\eta}, a_{w},b_{w}$ equal to 0.001; moreover we specify   $\mu_{\mu_{w}}=0$, $\sigma_{\mu_{w}}=10^4$. 
For the transition probabilities we specify an informative distribution: we set   $\alpha_{k,k}=0.5$ and $\alpha_{k,k-1}= \alpha_{k,k+1}=0.2$  for $k=2,\dots, K-1$, and spread the residual mass over the remaining elements ($\alpha_{k,j}= 0.1/(K-3), j=1, \dots, K; j\neq k $). For the extreme categories, $k=\{1, K\}$ we set $\alpha_{1,2}=0.2, \alpha_{1,j}= 0.2/(K-2)$ and $\alpha_{K,K-1}=0.2, \alpha_{K,j}=0.2/(K-2)$, respectively, $ j=1, \dots, K;  j\neq k$  ($k=\{1, K\}$, respectively).
This specification reflects our prior assumption that most of the observed categories are correctly specified and that the misclassification is more likely to {\blue involve} 
adjacent categories.
 To assess sensitivity of estimates to the prior specification,  we considered alternative non informative Dirichlet priors, as detailed in Section 1 of the Supplementary Materials. The results indicate that  inferences are robust to the choice of the vector $(\alpha_1,\dots ,\alpha_K)$, which can be ascribed to the high level in the hierarchy occupied by the Dirichlet prior.

For each model, we generate $10^{4}$ Markov chain Monte Carlo simulations, discarding the first half and then thinning  the chains by taking one out of every 10 sampled values. Chain convergence has been monitored by visual inspection using standard convergence diagnostic tools, such as trace plots and autocorrelation plots. 
\begin{figure}
\centering
\includegraphics[height=.47\textheight, width=.8\textwidth]{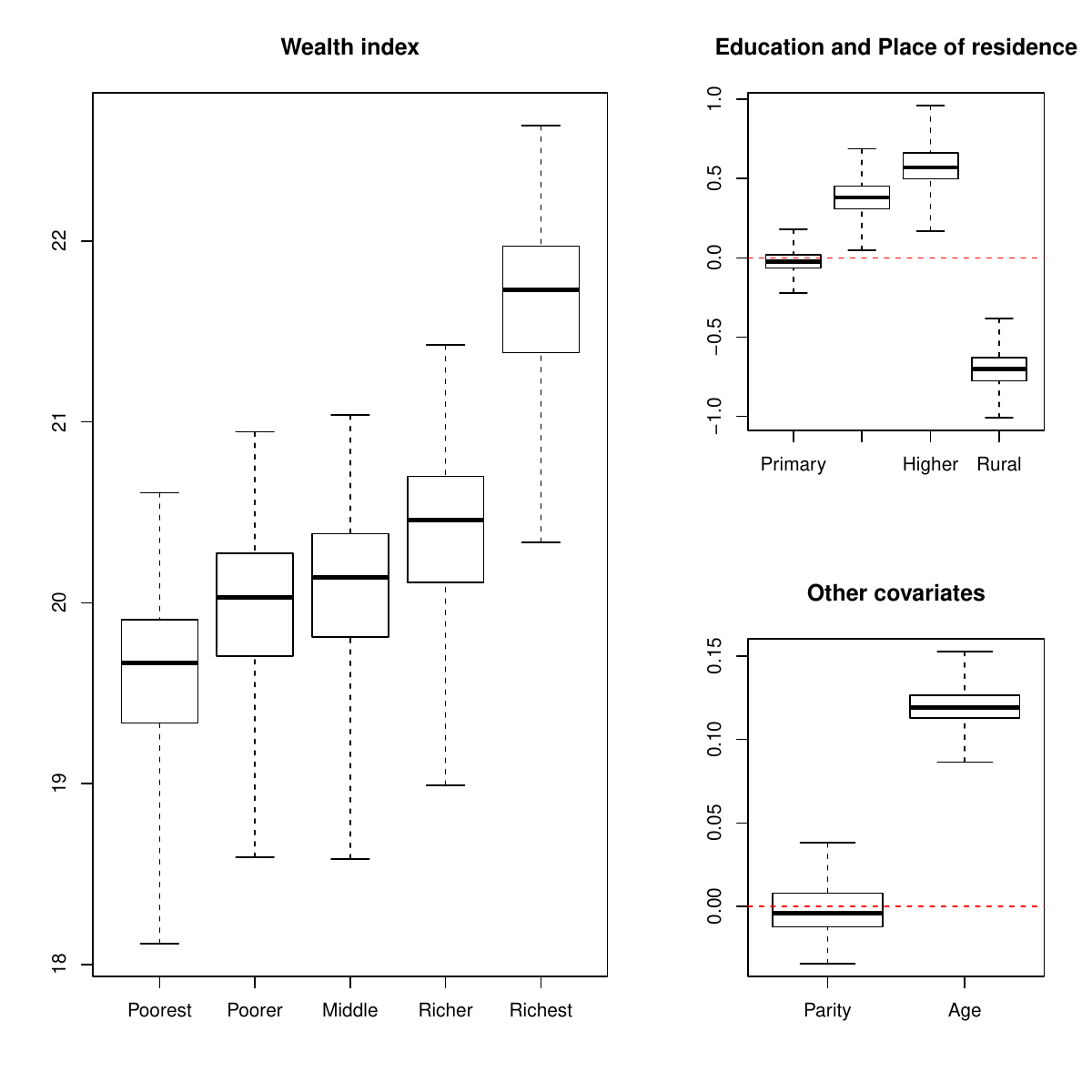}\vspace*{-1em}
\includegraphics[height=.47\textheight, width=.8\textwidth]{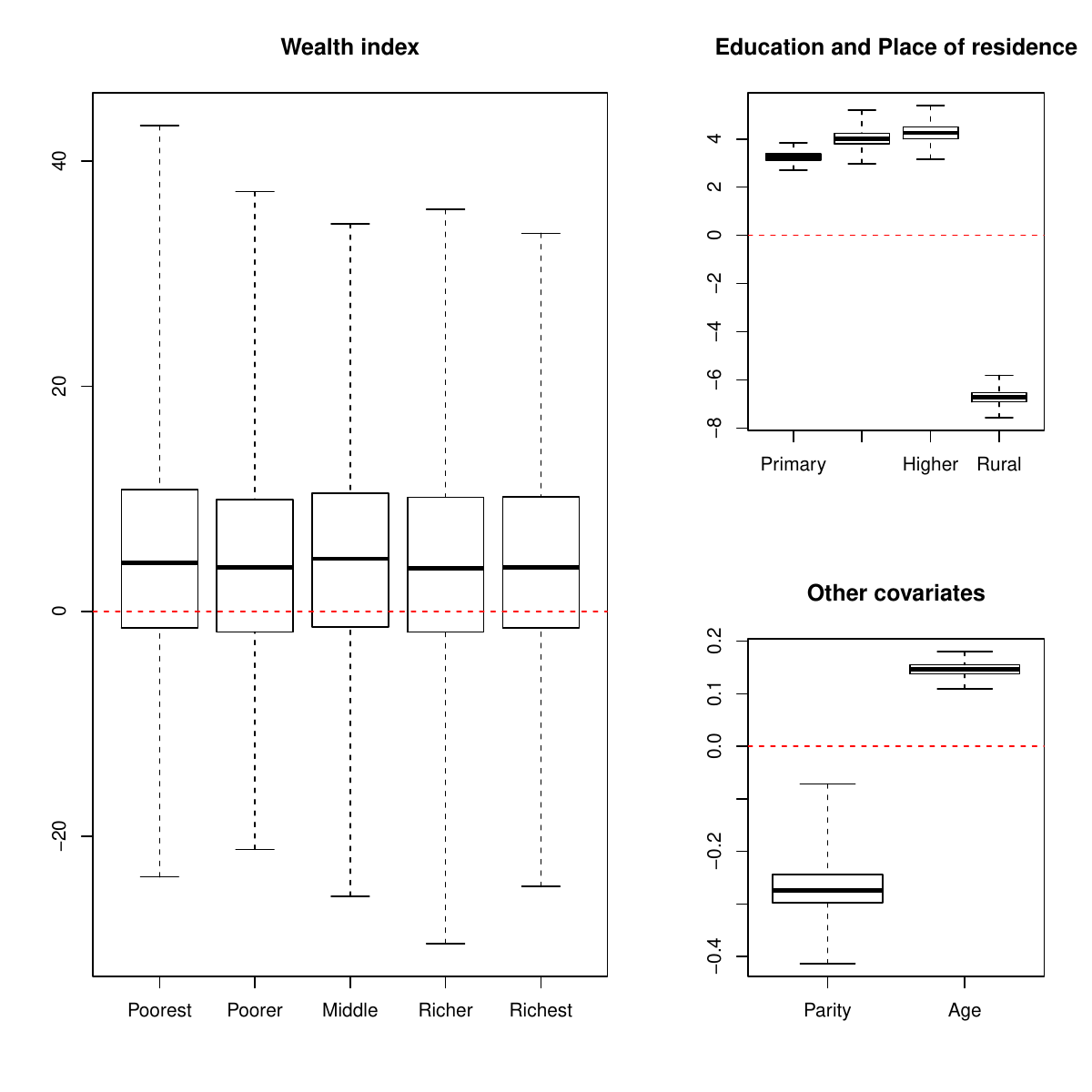}
\caption{Ethiopia data: posterior distributions of the model parameters under the proposed model (top panel) and under the assumption that the covariates are measured without error (bottom panel).
In the left panels, we show the posterior distribution of the parameter $\beta$ of the wealth index. In the right panels we present the posterior distribution of all the other regression parameters.}
\label{fig:01}
\end{figure}

We begin with an inspection of the estimates of the regression parameters obtained under the two models, with and without accounting for measurement error. Indeed, as highlighted by the simulation study, inaccurate estimation of regression coefficients may affect small area predictions. The  posterior distribution of the regression parameters under both models  are reported in  Figure~\ref{fig:01}. 
Under the measurement error model (top panel), the covariates' effects are all consistent with expectations. 
The BMI increases with the wealth index category, so that the poorest women are more likely to be underweight than the richest ones.  Although expected, such an important effect of the wealth index has not been always confirmed in previous studies. Also, education significantly impacts on the BMI, since more educated women show a larger BMI than less educated ones. The model also highlights the great disparity between urban and rural areas, where the women's undernutrition problem is more severe. 
The number of children ever born (parity) is another factor found to  affect women's nutritional status significantly. The BMI decreases with parity, which means that the risk of underweight increases with the number of children. With respect to age, the model highlights positive linear association with BMI:  younger women are more likely to be underweight than older ones. This fact is already documented in the literature and quite expected since adolescent women are at more at risk of problematic first birth (the mean age at first birth is 19.6), HIV infection, illegal abortion, all related to BMI through the health condition.\\ Noticeably, under the model that ignores the measurement error in wealth index and age, the strong differential effect of the wealth index disappears  (see the bottom panel of Figure~\ref{fig:01}). This is also consistent with findings in the literature, that sporadically identifies this variable's importance. With respect to the other parameters, while the meaning of the coefficients is coherent with those obtained with the proposed model, the variables' effect is considerably inflated. 

 Small area predictions have been computed as illustrated in Section~\ref{sec:03}, under a fully Bayesian approach and conditioning on the available information. This includes area-level population figures for the auxiliary variables measured without error, that we obtained from the Central Statistical Agency. Although for this application the observed data derive from a complex sampling design, the latter can be considered  noninformative \citep[][p.79]{Rao:2015} as the auxiliary variables in our model include the rural/urban classification, which is used in the sampling design. A non significant correlation coefficient  of 0.03114 between model's residuals and sampling weights further supports this conclusion.

\begin{figure}[bthp]
\centering
\centering \hspace*{-1em}\includegraphics[width=0.52\textwidth]{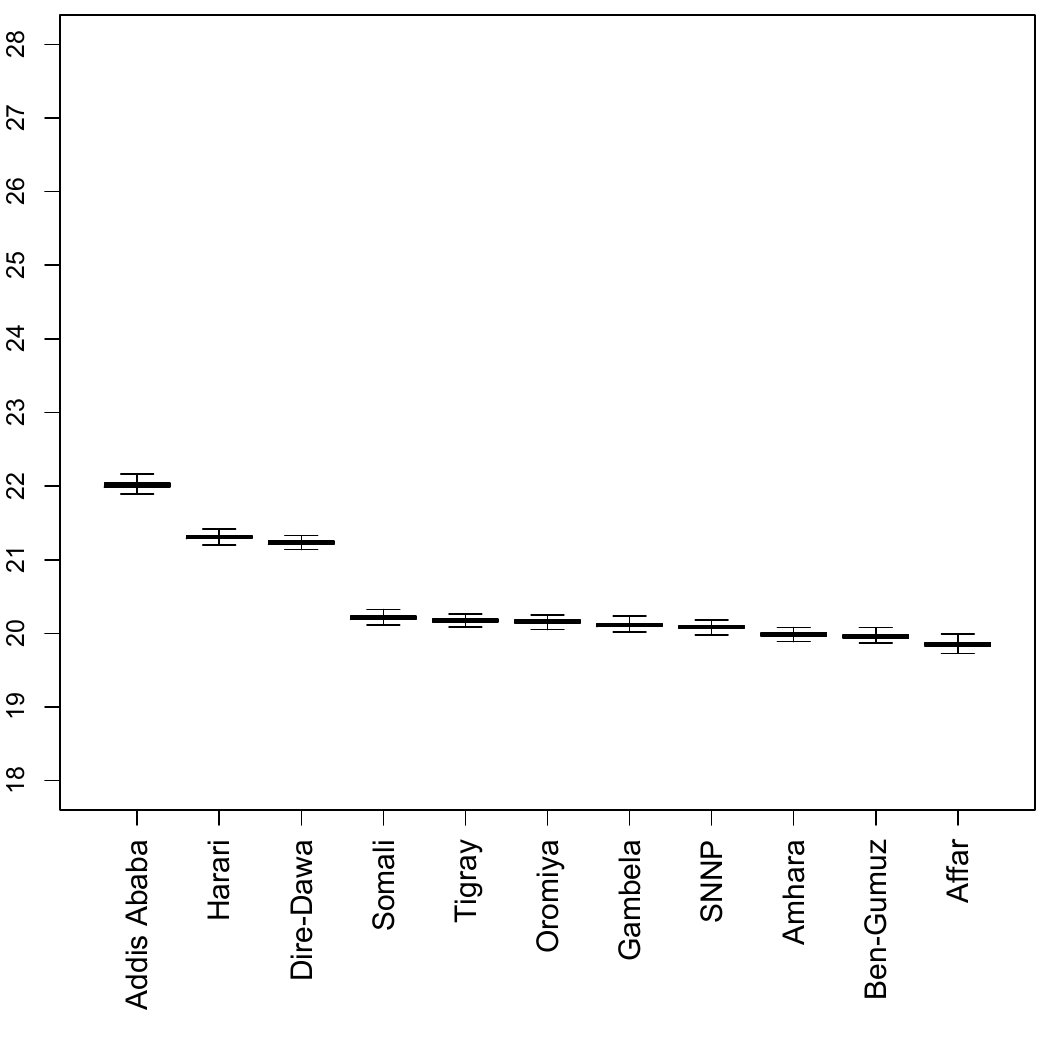}\hspace*{-0.2em}
\includegraphics[width=0.52\textwidth]{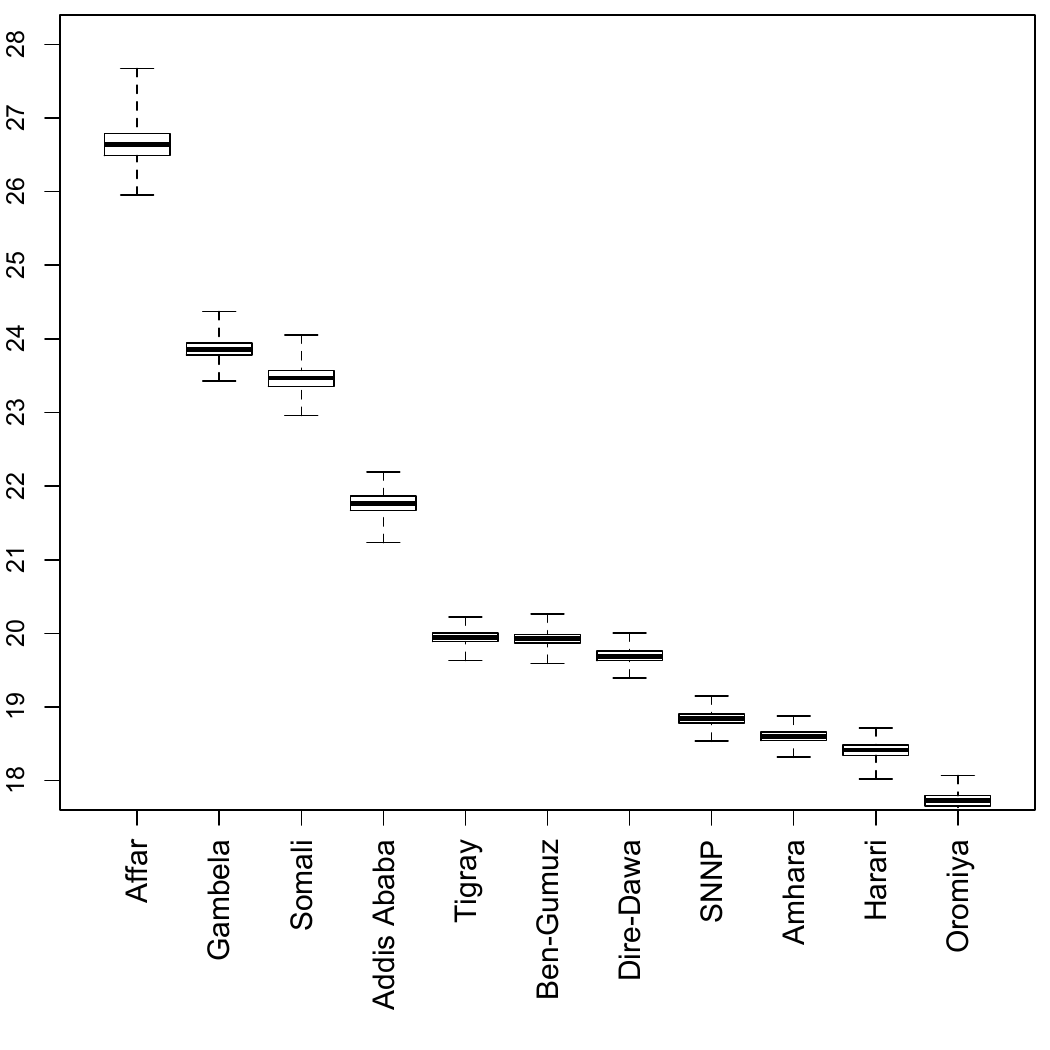}
\caption{Posterior distributions of the small area means for the Ethiopia DHS data. In the left panel, we show the estimates obtained with the proposed model; in the right panel, the small area means obtained with the model ignoring the measurement error.}
\label{fig:03}
\end{figure}

\begin{figure}[bthp]
\centering
\hspace*{-1em}\includegraphics[width=0.5\textwidth]{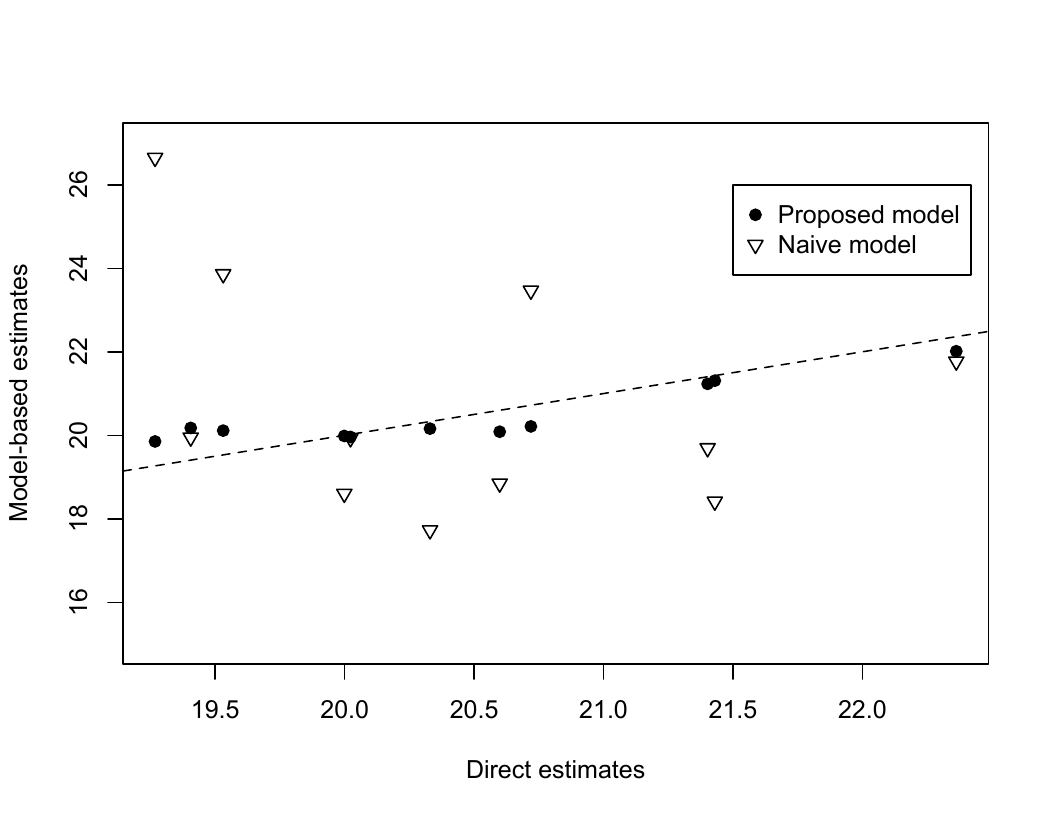} \hspace*{-0.2em}
\includegraphics[width=0.5\textwidth]{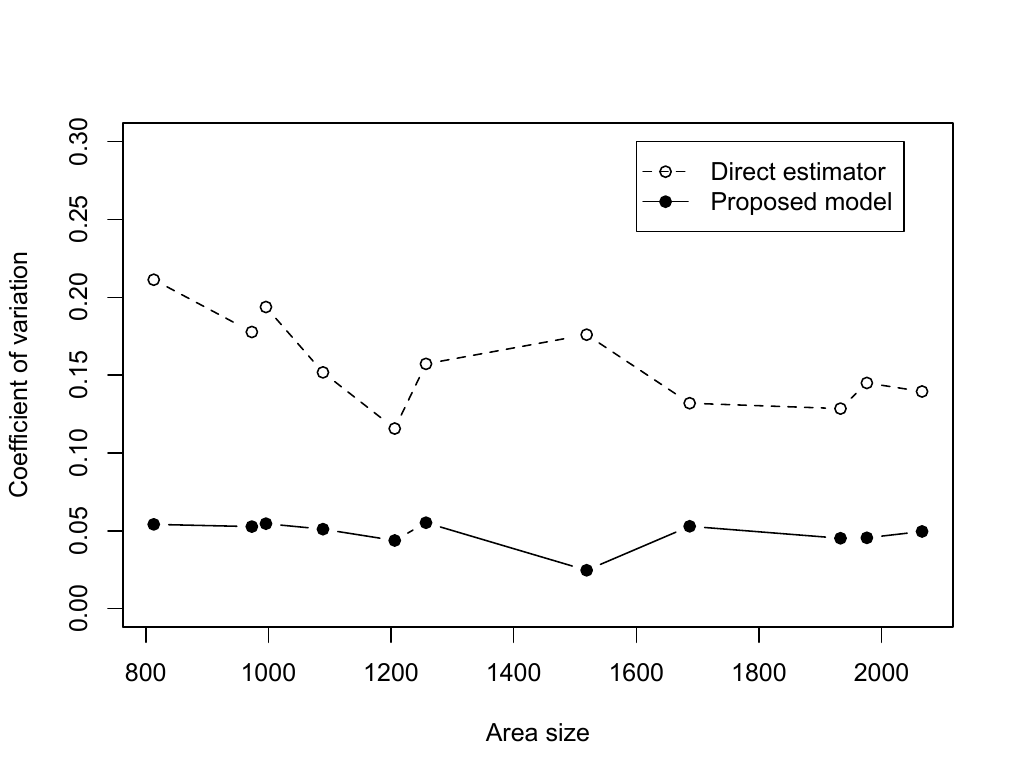}
\caption{Ethiopia DHS data: model-based small area mean estimates versus direct estimates (left panel) and CV plot (right panel)}
\label{fig:04}
\end{figure}

Given the strong impact of neglecting the measurement error in model estimation found in the application, and in light of the findings of the simulation study, we expect that small area predictions may differ considerably among models. Indeed in the simulation study the naive model leads to severely biased area estimates even when the perturbation level is low. The left panel of Figure~\ref{fig:03} shows the posterior distribution of the small area BMI means under the proposed model (left panel) and under the model that neglects the measurement error (right panel). Under the proposed model, the geographical pattern is coherent with expectations:  Addis Ababa, Harari and Dire Dawa show larger mean body mass index compared with all the other regions, that show very similar BMI distribution. Indeed, as noticed in \citet{EDHS},  more than one-third of women in Tigray, Affar, Somali, Gambela and Ben-Gumuz experience undernutrition.
On the other hand, small area means shown in the right of Figure~\ref{fig:03} do not reflect the geography, with even a reversal in the ranking of areas. Moreover, the latter estimates show a much stronger inter-area variation, resulting from inflation of model's parameters, that is not apparent under the proposed model. 
Although exhibiting a much reduced intra-regional variation, mean BMI area predictions under the proposed model are coherent with the direct estimates
and with the anticipated rural-urban disparities, as shown in Figure~\ref{fig:04}. The similarity between 
our small area estimates and the direct estimates can be ascribed to the fact that the sampling information is not negligible in this application; when the  area size is smaller (as in the simulation study in Table 3 of the Supplementary Materials), such similarity is not found.
Furthermore, comparison of the coefficients of variation of the small area predictions under the proposed model with those of the direct estimator indicates a great  reduction in the CVs, without introducing unnecessary shrinkage (see Figure~\ref{fig:04}).
We compare predictive performances of the proposed model and  of the model ignoring the measurement error according to  DIC and WAIC criteria \citep{Gelman}. For the  proposed model, $DIC=135001$ and  $WAIC= 162218.5$, while for the model ignoring the measurement error $DIC=269790.1$ and  $WAIC= 306275.2$; this confirms that the proposed model is preferable to the naive one also in predictive terms.
In conclusion, the application reveals that not accounting for measurement error may lead to misrepresentation of variable's effects and biased estimation of area-level figures, even in cases when the sample information is strong and the model-based predictions should agree with the direct estimates. On the other hand, as also testified by the use of predictive criteria, there is an advantage in introducing an appropriate small area model, and even in applications where the area size is not particularly small and the direct estimates might be considered reliable: indeed the small area estimates obtained with the proposed model are in strong agreement with the direct estimates, whereas those obtained neglecting the measurement error are very different.   

%
%
The proposed model allows for estimation of $P$. Table~\ref{tab7} shows  posterior mean and  posterior standard deviation of the misclassification probabilities for the wealth index. Notice that the transition matrix differs considerably from the prior specification and that changes are expected to occur essentially only for units with the lowest category of that variable. 
\begin{table}[t]
\centering
\begin{tabular}{cccccc}
  \hline
k & 1 & 2 & 3 & 4 & 5 \\ 
  \hline
1 & 0.223 (0.002)  & 0.154 (0.004) & 0.146 (0.003) & 0.122 (0.004) & 0.355 (0.023) \\ 
  2 & 0.055 (0.064) & 0.798 (0.100)  & 0.057 (0.107) & 0.051 (0.110) & 0.039 (0.103)\\ 
  3 & 0.032 (0.204) & 0.066 (0.120) & 0.778 (0.246) & 0.056 (0.117)  & 0.068 (0.113) \\ 
  4 & 0.048 (0.179) & 0.046 (0.126) & 0.040 (0.076) & 0.824 (0.306) & 0.042 (0.239) \\ 
  5 & 0.033 (0.088) & 0.030 (0.059) & 0.009 (0.104) & 0.008 (0.224) & 0.920 (0.134) \\ 
   \hline
 \end{tabular}
\caption{Ethiopia DHS data: posterior mean and standard deviation (in brackets) of the misclassification matrix $P$.}
\label{tab7}
\end{table}

\section{Discussion}
\label{sec:06}
In this paper, building on \citet{Ghosh:2006}, we have proposed a Bayesian unit-level measurement error small area model that accounts for  misclassification in categorical variables and allows for unknown perturbation mechanism. 
We investigated the performance of  the proposed model in estimating the regression parameters and  
predicting the small area means based on simulated and real data. We focused on Ethiopia DHS data to study the effect over women's body mass index of several social and demographic variables; random effects specified at the area level allow us to account for possible regional differences in the distribution of  BMI and permit accurate prediction of  women's mean BMI at the regional level. 
Other proposals in the literature, most notably MC-SIMEX \citep{Kuchenhoff:2006}, address the issue of misclassification in covariates. An important drawback of this method is that knowledge of $P$ is required. Excluding situations in which covariates are perturbed on purpose for confidentiality reasons, this assumption is often unrealistic also in a small area context. By contrast, our proposal offers the advantage to allow for unknown perturbation matrix $P$, that is estimated jointly with all the unknown model's parameters.

Based on simulated data, we also study  the 
model's capability in reconstructing the true value of the perturbed variables  for each unit. Under the assumption of unknown transition matrix $P$, not only is our model  able to reduce the estimation bias, but also to recover a large fraction of the original scores for the misclassified variable.  The proposed model also reveals a substantial robustness with respect to the prior distribution specifications.\\
Interestingly in our application, when measurement error is not accounted for by the model, the importance of the wealth index is masked by other variables' effect; the wealth index has been documented in the literature as a meaningful measure of socio-economic status, household food security and disposable income available for food \citep{corsi}. Moreover, the ranking of areas is not meaningful when error in covariates is neglected. 
Besides a strong effect of the socio-economic status, the results obtained under the proposed model reflect a noticeable rural-urban disparity and an effect of age on the BMI level. Once adjusting for individual and area-level covariates, regional disparities remain, but these are not as strong as the model that neglects measurement error would predict.

Our results indicate that variable weights and regression effects may be severely affected by neglecting  the presence of errors in covariates; we also conclude that  model selection may  be affected when measurement error in covariates is not properly accounted for by the model. Moreover, as evidenced both in the simulation study and in the application,  area predictions are subject to large RMSEs when measurement error is neglected{\blue; the corresponding increase in standard deviations indicates that area predictions are subject to large biases under the naive model},  which is a very important aspect in small area estimation. 
The application also shows that even in the presence of relatively large subsamples, a situation in which one could expect a reconciliation between the model based and the direct estimator, there is still an advantage in resorting to model based approach, provided measurement error is allowed for. This calls for a formal procedure to establish which covariates, if any, are measured with error: to our knowledge, this is still an open issue in the measurement error literature. Whereas practitioners are usually driven by their experience and knowledge of the survey, the problem can be cast within a  variable selection framework.  This will be the subject of future research.

Defining BMI as a categorical outcome (underweight, normal, overweight), \cite{corsi} used a Bayesian  multinomial logistic regression to  assess the socio-economic and geographic patterning of underweight and overweight in the population;  to investigate the existence of a neighbourhood effect they include random effects defined at the neighbourhood level. While the authors can address both issues of under- and over-nutrition, their model does not account for measurement error. Our model could also be extended to a generalized linear regression framework to allow for measurement error in covariates. Moreover, a large variation can be expected for the BMI within regions; at the same time, under the proposed model we observe clusters of areas with similar BMI means. This suggests that spatial effects could also be included in the model. These aspects will be the subject of  further research.

\bibliographystyle{ba}

\end{document}